\documentclass[twocolumn]{openjournal}

\usepackage{xcolor}
\usepackage{textgreek}
\usepackage{upgreek}
\usepackage[utf8]{inputenc}
\usepackage[english]{babel}
\usepackage{amsmath}

\usepackage{hyperref}
\hypersetup{
    unicode,
    colorlinks=true,
    linkcolor=linkcolor,
    citecolor=linkcolor,
    filecolor=linkcolor,
    urlcolor=linkcolor,
}
\usepackage{color,colortbl}
\definecolor{linkcolor}{rgb}{0.0,0.3,0.5}
\DeclareGraphicsExtensions{.bmp,.png,.jpg,.pdf}
\usepackage{verbatim}
\usepackage[normalem]{ulem}
\usepackage{orcidlink}
\usepackage{soul}

\graphicspath{ {./figs/} }

\begin{document}

\title{Discovery of a nearby ultra-fast magnetic white dwarf propeller:\\common pathways to magnetic cataclysmic variables}

\author{\vspace{-30pt}Antonio C. Rodriguez$^{1,2,3,\dagger}$ \orcidlink{0000-0003-4189-9668}}
\author{Kareem El-Badry$^{3}$ \orcidlink{0000-0002-6871-1752}}
\author{R. Michael Rich$^{4}$ \orcidlink{0000-0003-0427-8387}}
\author{Kevin N. Ortiz Ceballos$^{1}$ \orcidlink{0000-0003-3455-8814}}
\author{Ingrid Pelisoli$^{5}$ \orcidlink{0000-0003-4615-6556}}
\author{Casey J. Law$^{3}$ \orcidlink{0000-0002-4119-9963}}
\author{Kaya Mori$^{6}$ \orcidlink{0000-0002-9709-5389}}
\author{Boris Gaensicke$^{5}$ \orcidlink{0000-0002-2761-3005}}
\author{Edo Berger$^{1,2}$ \orcidlink{0000-0002-9392-9681}}
\author{Liam Connor$^{1}$ \orcidlink{0000-0002-7587-6352}}
\author{Ilkham Galiullin$^{7}$ \orcidlink{0000-0001-5778-2355}}
\author{Ilaria Caiazzo$^{8}$ \orcidlink{0000-0002-4770-5388}}
\author{Matthew Graham$^{3}$ \orcidlink{0000-0002-3168-0139}}
\author{Samuel Whitebook$^{3}$ \orcidlink{0000-0002-6836-181X}}

\email{Corresponding Author: antonio.c.rodriguez.1998@gmail.com}
\thanks{$\dagger$\; Independent Researcher, Mountain View, CA, USA}
\affiliation{$^1$Center for Astrophysics $|$ Harvard \& Smithsonian, 60 Garden Street, Cambridge, MA 02138, USA}
\affiliation{$^2$NSF Institute for AI and Fundamental Interactions (IAIFI), Cambridge, MA, USA}
\affiliation{$^3$Department of Astronomy, California Institute of Technology, 1200 E.\ California Blvd., Pasadena, CA 91125, USA}

\affiliation{$^4$Department of Physics and Astronomy, UCLA, 430 Portola Plaza, Box 951547, Los Angeles, CA, 90095-1547, USA}

\affiliation{$^5$Department of Physics, University of Warwick, Coventry CV4 7AL, UK}

\affiliation{$^6$Columbia Astrophysics Laboratory, Columbia University, New York, NY, USA}

\affiliation{$^7$Kazan Federal University, Kremlevskaya Str.18, 420008, Kazan, Russia}

\affiliation{$^8$Institute of Science and Technology Austria (ISTA), Am Campus 1, 3400 Klosterneuburg, Austria}

\begin{abstract}
The fastest spinning white dwarfs (WDs), which approach break-up speeds, are found in close binary systems. In the two known magnetic WD ``propellers'', AE~Aqr and LAMOST~J0240+1952, the magnetic WD spins so rapidly that it ejects the material transferred by its Roche-lobe-filling companion rather than accreting it. We report the discovery of the third such system, 1eRASS~J042633.5$-$250226 (1eRASS J0426/Valeria), at only 265 pc, identified by combining SRG/eROSITA and \textit{Gaia} data via the ``X-ray Main Sequence''. TESS photometry shows AE~Aqr-like flaring and orbital modulation with $P_{\mathrm{orb}} \approx 6.62$ hr, and ZTF shows month-long high/low state changes. Phase-resolved Palomar/DBSP spectroscopy confirms the orbital period and yields a donor radial velocity of $K_2 = 202\pm7$~km~s$^{-1}$. High-speed Palomar/CHIMERA photometry detects optical pulsations on three nights with the strongest detection at $P_\textrm{spin} = 26.8$ sec, though further observations are needed to confirm this as the WD spin. Spectral fitting of the donor in the low state reveals a late K-type star with $T_{\mathrm{donor}} = 4100\pm150$~K; the near Roche-filling condition then gives $M_{\mathrm{donor}} = 0.24\pm0.04~M_\odot$ and, through the mass function, $M_{\mathrm{WD}} = 0.51$--$0.86~M_\odot$ for $i = 90^{\circ}$--$50^{\circ}$. Zeeman splitting in the optical spectrum yields a candidate field strength of $B_{\mathrm{WD}} = 26 \pm 3$~MG, meaning the system will evolve into a polar. Like other WD propellers, the H$\alpha$ Doppler tomogram shows no disk or stream, and the system is X-ray faint ($L_X \approx 7\times 10^{30}$~erg~s$^{-1}$). Unlike the other two WD propellers, this system is not detected in the radio, and a spin period is not detected in X-rays by SRG/eROSITA. The proximity of this system and  evolutionary model constraints on the timeframe of the propeller phase yield a birth rate density of $\mathcal{R}\sim8\times10^{-16} \textrm{pc}^{-3} \textrm{yr}^{-1}$, which agrees with that of magnetic CVs, suggesting that a substantial fraction of magnetic CVs pass through a propeller phase.
\end{abstract}

\maketitle

\section{Introduction}
\label{sec:intro}
The spin of white dwarf (WD) stars is tied to their evolutionary history. Most isolated WDs spin with periods of 10--100 hours, an imprint of the angular-momentum and mass-loss processes of their progenitors \citep[e.g.][]{2003kawaler, 2017hermes}. WDs in close binaries, however, can be spun up dramatically by accretion. 

Cataclysmic variables (CVs) are semi-detached binaries in which a WD accretes from a Roche-lobe-filling companion \citep[e.g.][]{1995warner, 2001hellier}, and around 35\% of them host a magnetic WD whose field channels the accretion flow \citep{2020pala, 2025rodriguez_survey}. Angular momentum loss drives their evolution, causing CVs to evolve to shorter periods throughout their lifetimes, before reaching the canonical period minimum of $\approx$ 78 min. Magnetic CVs are critical for the study of WD rotation, since accretion onto the magnetic poles makes the spin directly observable. Younger magnetic CVs (most, but not all, intermediate polars; IPs; $P_{\mathrm{orb}} \gtrsim 3$~hr) tend to host rapidly spinning WDs, while older ones (most, but not all, polars; $P_{\mathrm{orb}} \approx78~\textrm{min} - 2$~hr\footnote{Recently, a metal-poor, old polar has been found at an even shorter orbital period of 64 min \citep{2025galiullin}.}) host WDs synchronised with the orbit. This motivates the picture that high accretion rates early in CV evolution spin the WD up toward break-up ($P_{\mathrm{spin}}=20$--$30$~s), powering an internal dynamo that magnetises the WD to $10^{6}$--$10^{8}$~G \citep{2021schreiber, 2022ginzburg}. The internal WD dynamo has been hypothesized to be a convective dynamo, similar to the Earth's, operating due to the crystallization of the carbon-oxygen interior. Recent surveys have supported this, showing that isolated white dwarfs become more magnetic as they age, though the role of crystallization is still under debate \citep{2022bagnulo, 2026castro}.

Close magnetic WD binaries span a rich phenomenology: intermediate polars (IPs; 1--10~MG fields, freely rotating WDs with $P_{\mathrm{spin}} \approx 30$--3000~s) \citep{1994patterson}, polars (10--240~MG, spin synchronised with the orbit) \citep[e.g.][]{1990cropper}, and asynchronous polars \citep[e.g.][]{2014boyd} make up the main categories of magnetic CVs. Other exotic close magnetic WD binaries that have been hypothesized to explain intermediate evolutionary phases include: radio-pulsing ``WD pulsars'' \citep{2016marsh, 2023j1912, 2025castro_segura}, hybrid systems sharing properties of both WD pulsars and polars \citep{2024discovery}, and most recently, WDs in long-period radio transients \citep[][]{2025rodriguez, 2025deruiter, 2026rodriguez, 2026rose, 2026mo}.

Here we focus on ``WD propellers'', which host the fastest-spinning WDs known. Notably, some IPs and even non-magnetic CVs have shown to be rapidly spinning, approaching 30-sec spin periods. \cite{2024salcedo} features a review of such systems. Though sometimes labelled a subset of IPs, WD propellers are not accreting: the WD spins so fast that magnetic pressure overcomes ram pressure and matter is ejected from the system rather than accreted. Only two are known: AE~Aqr \citep[$d=91$ pc, $P_{\mathrm{spin}}=33.1$~s,  $P_{\mathrm{orb}}=9.88$~hr;][]{1979wynn_aeaqr, 1979crampton, 1997wynn} and LAMOST J024048.51+195226.9 \citep[$d=620$ pc, $P_{\mathrm{spin}}=24.93$~s, $P_{\mathrm{orb}}=7.34$~hr;][]{2020thorstensen, 2021garnavich, 2021pretorius, 2025pelisoli_j0240}. Both share long binary periods, warm K-type Roche-lobe-filling donors, low X-ray luminosities, and high-amplitude optical flaring linked to the magnetic ejection of mass.

In this paper we present the discovery of 1eRASS~J042633.5$-$250226 (hereafter 1eRASS~J0426 or Valeria\footnote{We have chosen this as a common name for this system due to the ``V''-shaped asterism in the finder chart, and for the ancient Roman name meaning ``strength''. We note that the asynchronous polar ``Paloma'' was named in a similar manner \citep{2007Schwarz}.}), a third magnetic WD propeller at only $265.4 \pm 1.7$~pc, just over four times the distance to the nearest magnetic CV. Section~\ref{sec:discovery} describes the discovery and archival photometry; Section~\ref{sec:data} the multi-wavelength follow-up; Section~\ref{sec:results} the binary parameters; and Section~\ref{sec:discussion} places 1eRASS~J0426 in the context of the WD-propeller population and magnetic-CV evolution.

\section{Discovery}
\label{sec:discovery}

\begin{figure}
    \centering
    \includegraphics[width=0.95\linewidth]{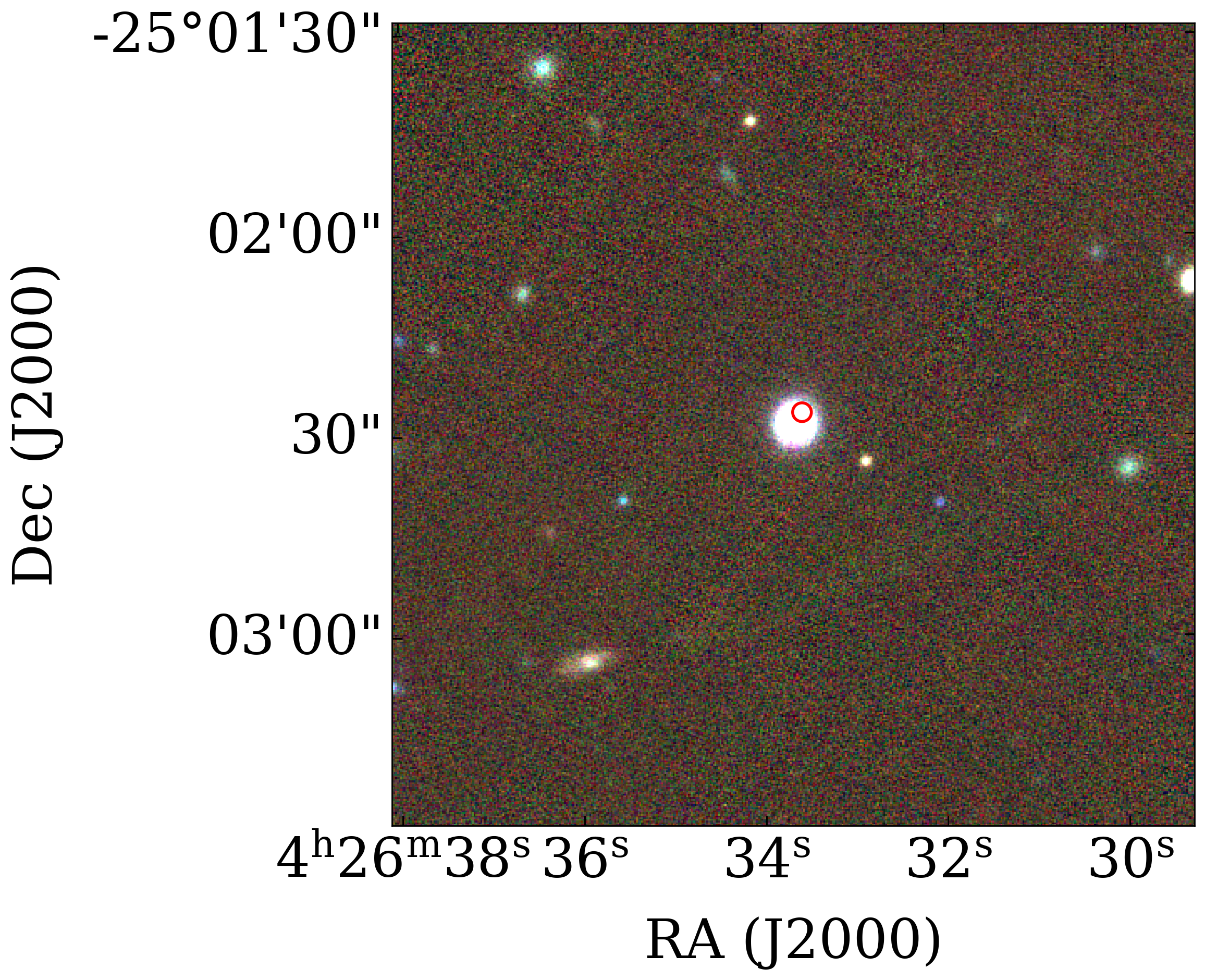}
    \caption{Optical multi-colour image of 1eRASS~J0426/Valeria (centre) with the SRG/eROSITA X-ray position (red; positional error 1.4~arcsec). The chance-alignment probability is $\sim 2.5$\%.}
    \label{fig:finder}
\end{figure}

\begin{figure*}
    \centering
    \includegraphics[width=\textwidth]{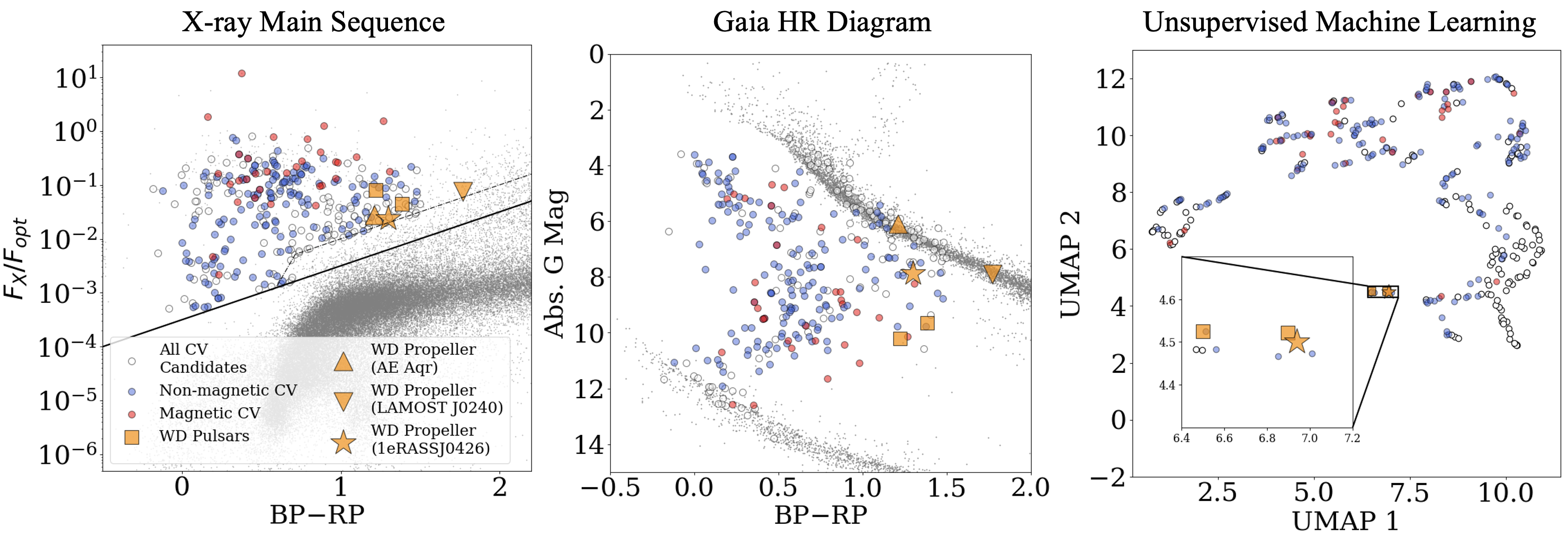}
    \caption{1eRASS~J0426/Valeria was identified via three independent techniques: the ``X-ray Main Sequence'' \citep[\textit{left};][]{2024rodriguez, 2025rodriguez_survey}, the \textit{Gaia} HR diagram (\textit{centre}), and unsupervised machine learning using UMAP on combined X-ray/optical data including \textit{Gaia} XP spectra (\textit{right}). In all three, the sources nearest known WD pulsars in the 1-kpc eRASS1 + \textit{Gaia} crossmatch were targeted. The marker for LAMOST J0240 on the left panel is an upper limit, since an X-ray detection has not been reported in the literature.}
    \label{fig:discovery}
\end{figure*}

This system was discovered in the 1-kpc volume-limited survey of CV candidates in SRG/eROSITA eRASS1 \citep{2025rodriguez_survey}, based on the ``X-ray Main Sequence'' \citep{2024rodriguez}, which crossmatched the January 2024 Western-Galactic-Hemisphere release of eROSITA aboard SRG \citep{2021sunyaev, 2021erosita, 2024merloni} against \textit{Gaia} DR3. Figure~\ref{fig:finder} shows the optical image and X-ray localization; the chance-alignment probability for an X-ray source of this brightness within $r_{\mathrm{sep}}=2.69$~arcsec of the optical counterpart is $\approx 2.5$\%. Figure~\ref{fig:discovery} shows the three parameter spaces in which 1eRASS~J0426 was flagged, in each case lying close to known WD pulsars. The first two are ``traditional'' approaches, relying on physics-motivated parameter spaces. The third is an unsupervised machine learning based approach, UMAP, which maps objects from a high parameter space to a lower parameter space in which distance serves as a proxy for similarity \citep{mcinnes2018umap}. In our case, the initial parameters fed into the UMAP algorithm were the \textit{Gaia} absolute magnitudes, X-ray luminosity, and normalized coefficients of the \textit{Gaia} XP spectra \citep{DeAngeli2023}. AE Aqr and LAMOST J0240 are also shown as references, though the latter is shown as an upper limit in the left-hand plot as no X-ray detection is reported in the literature\footnote{An analysis of the publicly available Swfit/XRT data (ObsID: 00013899001) rendered a non-detection, with an upper limit of $4.5\times10^{-13} \textrm{erg s}^{-1}\textrm{cm}^{-2}$ in the 0.2--10 keV range, assuming a power law index of 2. }

\subsection{Archival Photometry}
\label{sec:archival}

\begin{figure*}
    \centering
    \includegraphics[width=0.8\linewidth]{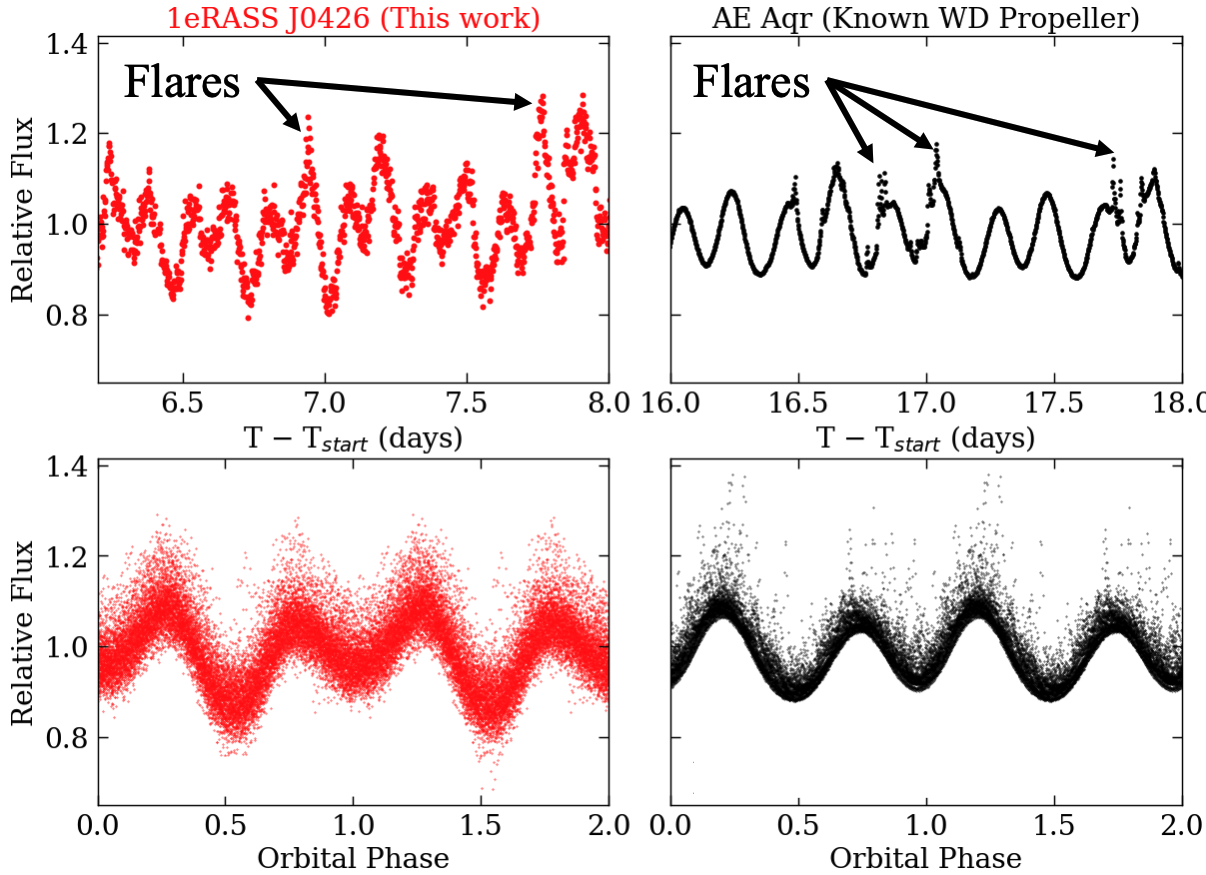}
    \caption{TESS light curve of 1eRASS~J0426/Valeria (\textit{black}) folded on the 6.616-hr orbital period, showing ellipsoidal orbital variability superposed with high-amplitude flares similar in morphology to AE~Aqr (9.88-hr orbital period).}
    \label{fig:tess}
\end{figure*}

We searched the archive of the Transiting Exoplanet Survey Satellite \citep[TESS;][]{2015tess} using \texttt{lightkurve} \citep{2018lightkurve} and found that 1eRASS~J0426 bears a striking similarity to AE~Aqr, the prototypical magnetic WD propeller: both show orbital modulation superposed with large-amplitude flares from ejected material (Figure~\ref{fig:tess}). A Lomb--Scargle periodogram of the flare-clipped TESS light curve peaks at half the orbital period (Appendix~\ref{sec:appendix_tess}). A three-harmonic fit to the flare-clipped light curve with a free period yields the photometric ephemeris adopted throughout: $P_{\mathrm{orb}} = 6.61612 \pm 0.00010$~hr, with the (deeper) photometric minimum at $T_0(\mathrm{BJD_{TDB}}) = 2459144.6375 \pm 0.0002$. Long-baseline photometry from the Zwicky Transient Facility \citep[ZTF; Figure 2; ][]{2019bellm} shows transitions between high and low photometric states on $\sim$month timescales, which we interpret as state changes in the propeller activity, mirroring AE~Aqr's low/high flaring states \citep[e.g.][]{1979wynn_aeaqr}.

\begin{figure*}
    \centering
    \includegraphics[width=0.92\linewidth]{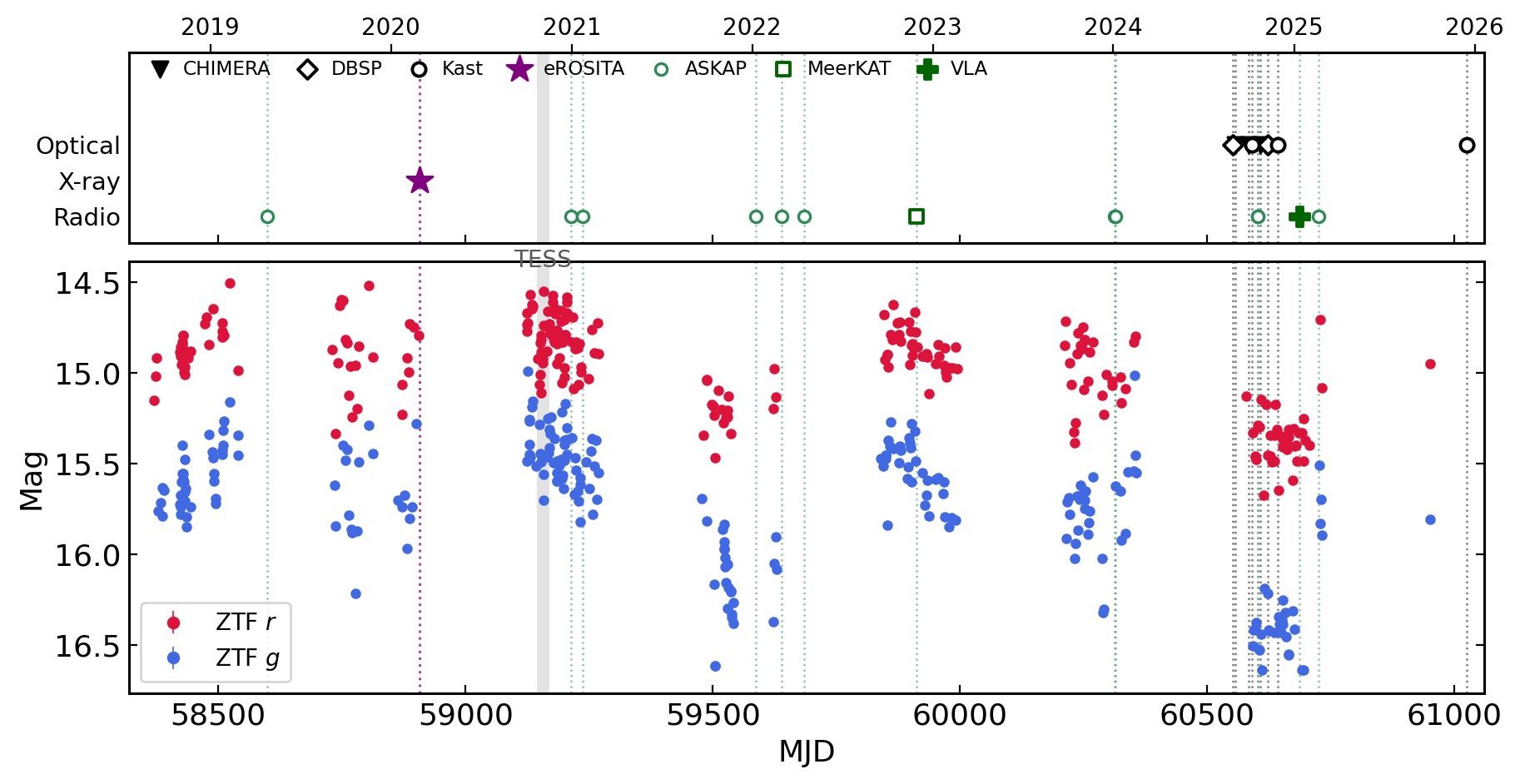}
    \caption{Long-baseline ZTF $g$- and $r$-band photometry of Valeria (\textit{bottom panel}), with the epochs of all follow-up and archival observations marked (\textit{top panel}): optical spectroscopy (Palomar/DBSP, open diamonds, 2024 Aug 30 and Nov 9; Lick/Kast, open circles, 2024 Oct 7, 2024 Nov 28, 2025 Dec 16) and high-speed photometry (Palomar/CHIMERA, filled triangles; 2024 Sep 3, Sep 30, Oct 24); X-ray (SRG/eROSITA eRASS1 scan, star; 2020 Feb 28); and radio (ASKAP archival epochs, open circles; MeerKAT archival epoch, open square; VLA, filled cross; 2026 Jan 6). Dotted vertical lines extend each epoch down through the light curve. The TESS Sector 31 window (2020 Oct 22 -- Nov 16) is shaded grey. The system transitions between an active (flaring) high state and a quiescent low state on $\sim$few month- to year-long timescales.}
    \label{fig:ztf}
\end{figure*}

\section{Multiwavelength Follow-up}
\label{sec:data}

\subsection{Optical High-Speed Photometry}
\label{sec:chimera}

\begin{figure*}
    \centering
    \includegraphics[width=0.85\linewidth]{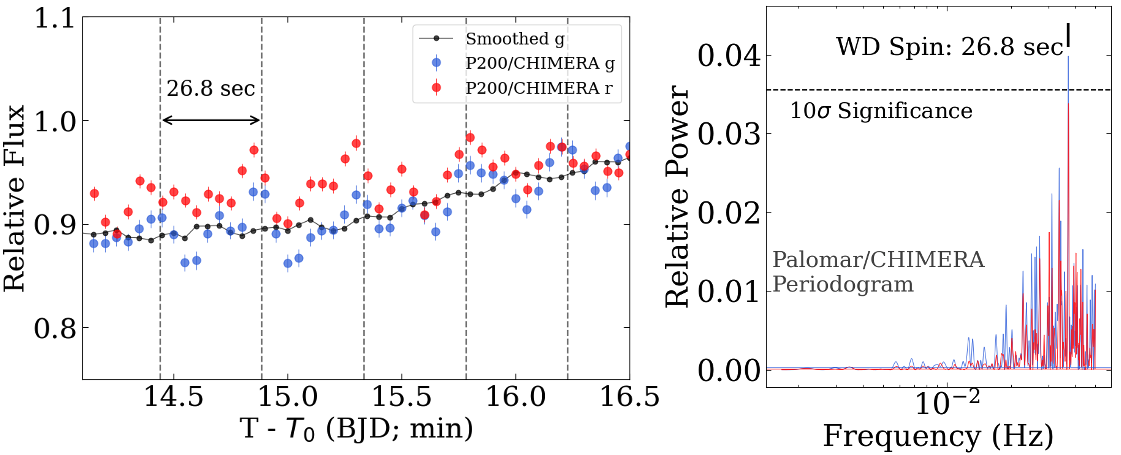}
    \caption{Palomar/CHIMERA high-speed (3~s) optical photometry of Valeria from 2024 September 3, the night with the strongest pulsation signal, showing a coherent $\approx 26.8$~s pulsation. The candidate period varies between nights (26--30~s on Sep 30; 19--24~s on Oct 24; Appendix~\ref{sec:appendix_chimera}), and the true spin period may be twice the pulse period if both magnetic poles contribute; all candidates satisfy $P_{\mathrm{spin}} < 60$~s (Section~\ref{sec:chimera}).}
    \label{fig:chimera}
\end{figure*}

We obtained 3-second cadence photometry of 1eRASS~J0426 with CHIMERA \citep{2017buckley_chimera} on the Palomar 200-inch telescope on three nights: 2024 September 3, September 30, and October 24 (UT). Each night reveals coherent optical pulsations at short periods, but the candidate period differs between nights. On September 3 -- the night with the most erratic flaring and by far the strongest pulsation signal -- the periodogram is best fit by 26.8~s, with a nearly equally strong alias at 26.7~s (Figure~\ref{fig:chimera}). On September 30 the blue and red channels favour $\approx$30~s and 26~s respectively, and on October 24 the strongest candidates are 19~s and 24~s (Appendix~\ref{sec:appendix_chimera}). Such behaviour is familiar from AE~Aqr, in which the optical pulse amplitude and phase wander with the flaring activity, and in which the optical signal is dominated by the \emph{first harmonic} of the spin because both magnetic poles contribute; the true spin period of 1eRASS~J0426 may therefore also be twice the observed pulse periods ($\sim$40--60~s).

We consequently treat the spin period as not yet uniquely determined. Two robust statements hold: (i) coherent short-period optical pulsations are present on all three nights, and (ii) all candidate periods -- including the possible factor-of-two ambiguity -- likely satisfy $P_{\mathrm{spin}} < 60$~s. Throughout this paper we adopt $P_{\mathrm{spin}} < 60$~s as the conservative constraint and refer to 26.8~s, the strongest single detection, as the leading candidate. Nevertheless, it could still be the case that this putative periodicity is due to flaring; the biggest red flag at the moment is the lack of agreement between the strongest periods in the different CHIMERA photometric channels. A dedicated multi-night high-speed campaign is required to phase-connect the pulsations and determine the spin period uniquely.

\subsection{Optical Spectroscopy}
\label{sec:spec}

\begin{figure*}
    \centering
    \includegraphics[width=0.64\linewidth]{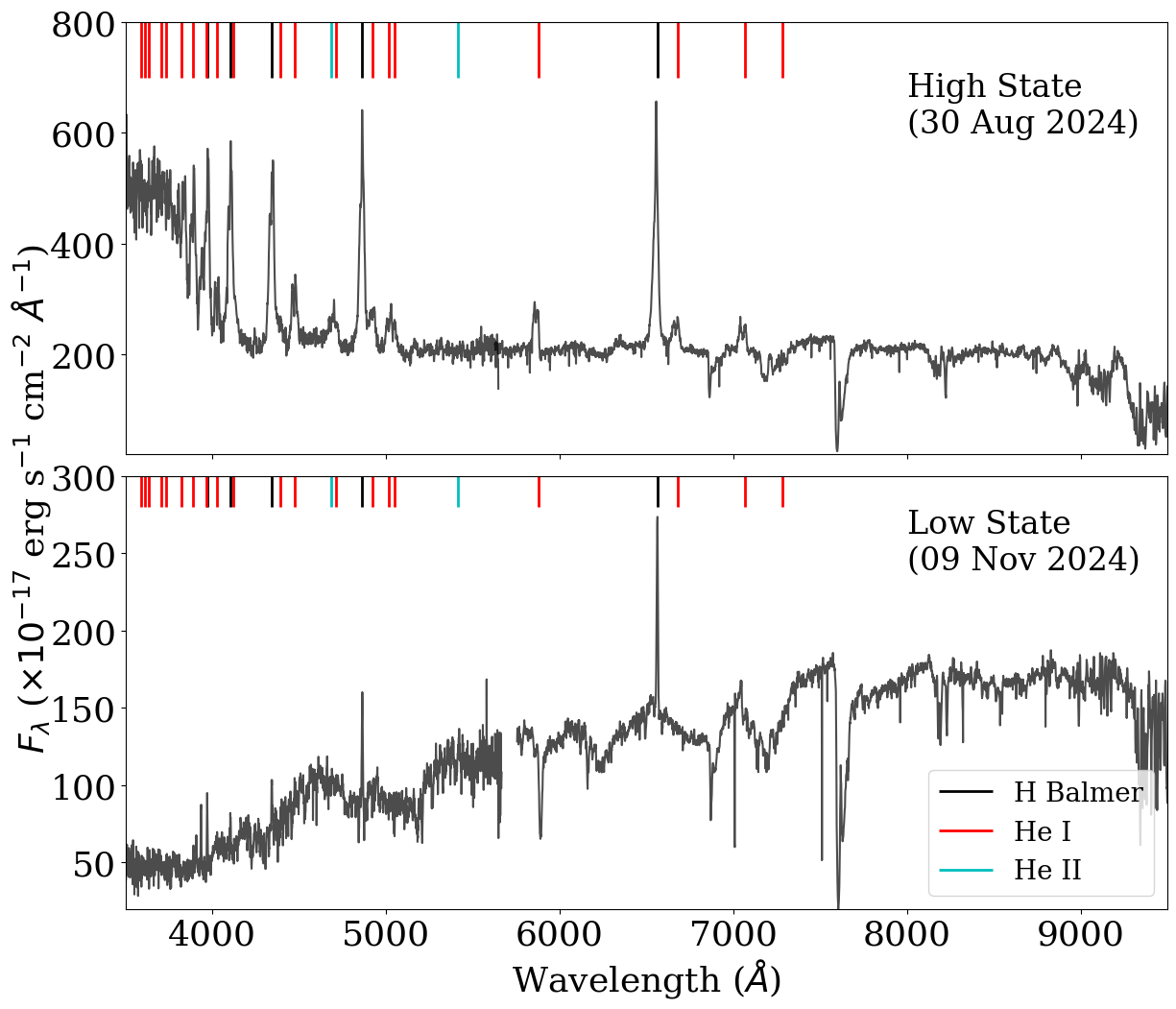}\includegraphics[width=0.36\linewidth]{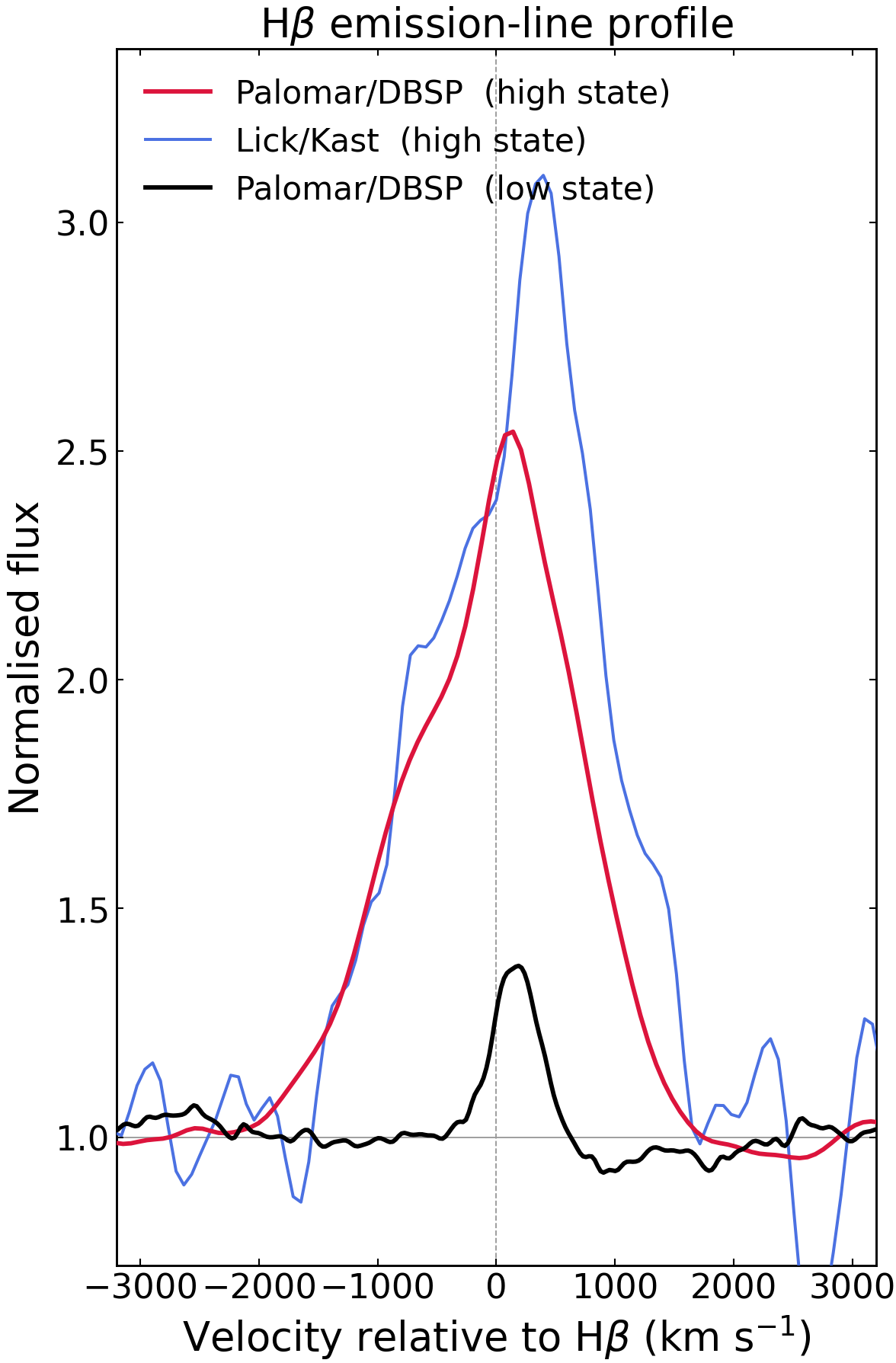}
    \caption{\textit{Left:} Palomar/DBSP spectra of Valeria in high (\textit{top}; 30 August 2024) and low (\textit{bottom}; 9 November 2024) states. The high state shows strong Balmer, He~I, and He~II~$\lambda$4686 emission; the low state is donor-dominated with only weak chromospheric H$\alpha$ emission. \textit{Right:} High and low state profiles of the H$\beta$ line reveal high speed ($\sim2000$ km/s) wings from ejected material and narrow ($\sim200$ km/s) chromospheric emission, respectively.}
    \label{fig:spec}
\end{figure*}

Phase-resolved optical spectroscopy was obtained with the Double Spectrograph (DBSP) on the Palomar 200-inch telescope on 30 August 2024 UT (high state) and 9 November 2024 UT (low state). The high-state spectrum shows strong, broad H, He~I, and He~II emission resembling that of magnetic CVs; the low-state spectrum is donor-dominated (TiO bandheads, the Na~I 8183/8195 doublet, and the Ca~II infrared triplet) with only weak H$\alpha$ emission (Figure~\ref{fig:spec}). A series of $38\times600$~s exposures spanning a complete orbit on 9 November is used for the radial-velocity analysis in Section~\ref{sec:rv}. Additional low-resolution spectra were obtained with the Kast spectrograph on the Shane 3-m telescope at Lick Observatory on 2024 October 7, 2024 November 28, and 2025 December 16 (UT); they are presented in Appendix~\ref{sec:appendix_kast}.

\subsection{X-ray}
\label{sec:xray}

1eRASS~J0426 was detected in the eRASS1 all-sky survey (which scanned this field on 2020 February 27--28) at a 0.2--2.3~keV flux of $F_X \approx 7\times 10^{-13}$~erg~s$^{-1}$~cm$^{-2}$ \citep{2024merloni}, corresponding to $L_X \approx 6\times 10^{30}$~erg~s$^{-1}$ at 265.4~pc -- two to three orders of magnitude below typical magnetic CVs, and consistent with AE~Aqr ($L_X\sim 10^{31}$~erg~s$^{-1}$) and (the upper limit of) LAMOST~J0240, supporting the picture that very little material reaches the WD.

An absorbed thermal-bremsstrahlung fit to the eROSITA spectrum (135 photons across all telescope modules) gives $kT \approx 10$~keV -- poorly constrained from above, with $kT \gtrsim 2$~keV robust -- and $N_H \approx 6\times 10^{20}$~cm$^{-2}$, near the Galactic value. The hardness ratio, $HR_1 = 0.81 \pm 0.06$, places the source on the hard-spectrum locus of magnetic CVs, with no soft blackbody excess and with a response-folded AE~Aqr-like plasma \citep{2008mauche} only slightly softer. The spectral analysis and figures are presented in Appendix~\ref{sec:appendix_xray}.

We searched the 113 in-band eROSITA photons -- spread over nine $\sim$30~s scan passes spaced 4~hr apart -- for periodic modulation at the candidate spin periods. After barycentring the photon arrival times (the barycentric delay drifts by 0.3 spin cycles across the 32-hr baseline), the Rayleigh power at 26.8~s rises to $Z^2_1 = 7.5$, a marginal $\sim$2$\sigma$ (window-corrected) enhancement at the a-priori optical period (Figure~\ref{fig:xray_period} in Appendix~\ref{sec:appendix_xray}) -- suggestive, but not a detection. The 99\% upper limit on the X-ray pulsed fraction at the spin period is $\approx 40$\%. 

\subsection{Radio Campaign}
\label{sec:radio}

\begin{figure}[h!]
    \centering
    \includegraphics[width=0.48\textwidth]{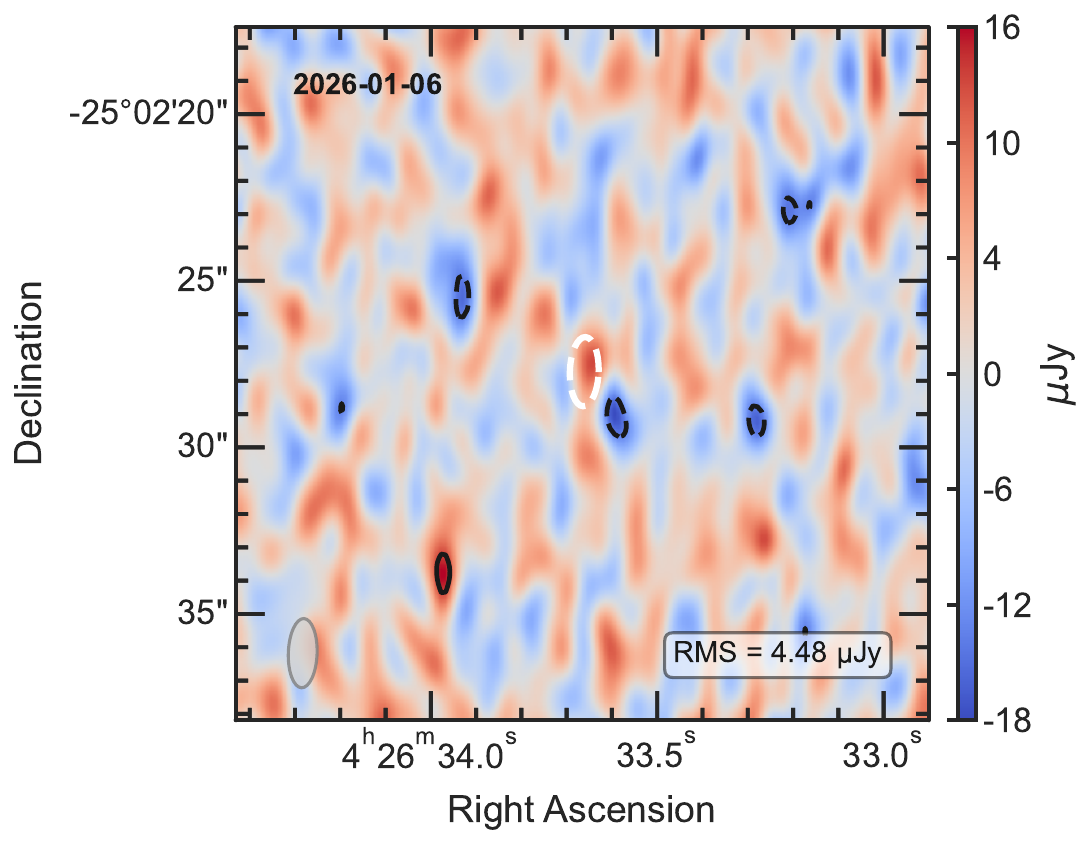}
    \caption{VLA C-band image of the field of Valeria. The synthesized beam is shown in the bottom left corner. The outlined beam in the centre of the image marks the \textit{Gaia} proper-motion-corrected coordinates of the Gaia DR3 counterpart at the observation epoch. Contours trace $-3\sigma$, 3$\sigma$, 5$\sigma$, and 7$\sigma$.}
    \label{fig:vla}
\end{figure}

We observed the field of 1eRASS~J0426 with the Karl G. Jansky Very Large Array (VLA) on 2026 January 6 in C-band (4--8~GHz) under Director's Discretionary Time program VLA/25A-452 (PI: Rodriguez). We observed a single 1-hour epoch in standard continuum mode, with 3C147 as the bandpass calibrator and J0457$-$2324 as the phase calibrator. The observation was processed and calibrated with the standard NRAO VLA pipeline (v2025.1.0.32), included in the Common Astronomy Software Applications \citep[CASA v6.6.6.17;][]{casa_team_casa_2022}, and image products were generated through the NRAO Science Reference Data Products (SRDP) program. We retrieved the products from the NRAO archive and use the self-calibrated image product for our science measurement. The resulting image is shown in Figure~\ref{fig:vla}. We do not detect any source at the position of 1eRASS~J0426, and establish a 3$\sigma$ upper limit of 13.4~$\upmu$Jy. The observation took place during a low (propeller-off) state (Figure~\ref{fig:ztf}).

We also searched the MeerKAT archive for archival observations that included the position of 1eRASS~J0426 in their field of view. We found one observation (Capture block ID: 1669756085, PI: G.~Di~Gennaro) taken on 2022 November 29--30 in the standard L-band (0.8--1.7~GHz) continuum observing mode, with 4.86 hours on-source. 1eRASS~J0426 was $\sim 0.2^{\circ}$ offset from the pointing centre, well within the $\sim$45~arcmin primary beam. We used the \texttt{OXKAT} pipeline \citep{heywood_oxkat_2020} to image the calibrated measurement set obtained from the MeerKAT archive, and find no significant source at the position of 1eRASS~J0426.

We also searched CASDA for archival ASKAP observations that cover the position of 1eRASS~J0426. We found multiple epochs of the Rapid ASKAP Continuum Survey (RACS) in all three bands (low-, mid-, and high-band), as well as an epoch from the FLASH survey. The source was not detected in any ASKAP epoch; the resulting upper limits are listed in Table~\ref{tab:radio}.

\begin{table}[ht!]
\caption{ASKAP flux limits for 1eRASS~J0426. \label{tab:radio}}
\centering
\small
\begin{tabular}{llll}
\hline\hline
Band/Project & SBID & Epoch & $S_{\mathrm{lim}}$ (mJy) \\
\hline
FLASH Pilot & 37797 & 2022-03-02 & $<0.24$ \\
RACS-high   & 35539 & 2022-01-08 & $<0.60$ \\
RACS-low    &  8579 & 2019-04-26 & $<1.02$ \\
RACS-low    & 39505 & 2022-04-16 & $<0.43$ \\
RACS-low    & 56711 & 2024-01-04 & $<1.74$ \\
RACS-low    & 56712 & 2024-01-04 & $<2.51$ \\
RACS-low    & 56713 & 2024-01-04 & $<0.54$ \\
RACS-low    & 56765 & 2024-01-05 & $<1.20$ \\
RACS-mid    & 20574 & 2020-12-31 & $<3.88$ \\
RACS-mid    & 21830 & 2021-01-23 & $<3.17$ \\
RACS-mid    & 21831 & 2021-01-23 & $<0.49$ \\
RACS-mid    & 66829 & 2024-10-19 & $<3.62$ \\
RACS-mid    & 66869 & 2024-10-20 & $<0.58$ \\
RACS-mid    & 71123 & 2025-02-20 & $<3.39$ \\
\hline
\end{tabular}
\end{table}

\section{Binary Parameters}
\label{sec:results}

We constrain the binary with three independent measurements: (i) the donor radial-velocity solution, which sets the mass function; (ii) a multi-band SED fit, which pins the donor temperature and radius; and (iii) the spin constraint $P_{\mathrm{spin}} < 60$~s, which sets a break-up lower limit on $M_{\mathrm{WD}}$.

\subsection{Radial-Velocity Analysis}
\label{sec:rv}

We extract donor RVs from the low-state DBSP red-arm time series by cross-correlating each continuum-normalised spectrum with a smoothed $T_{\mathrm{eff}}=3500$~K BT-Settl/PHOENIX template \citep{2011allard}, jointly over three telluric-clean, line-rich windows: the Na~I 8183/8195 doublet, the TiO bandhead complex at 7000--7150~\AA{}, and the Ca~II infrared triplet. Per-spectrum uncertainties come from the CCF curvature, and all velocities are corrected to the heliocentric frame.

\begin{figure*}
    \centering
    \includegraphics[width=0.97\linewidth]{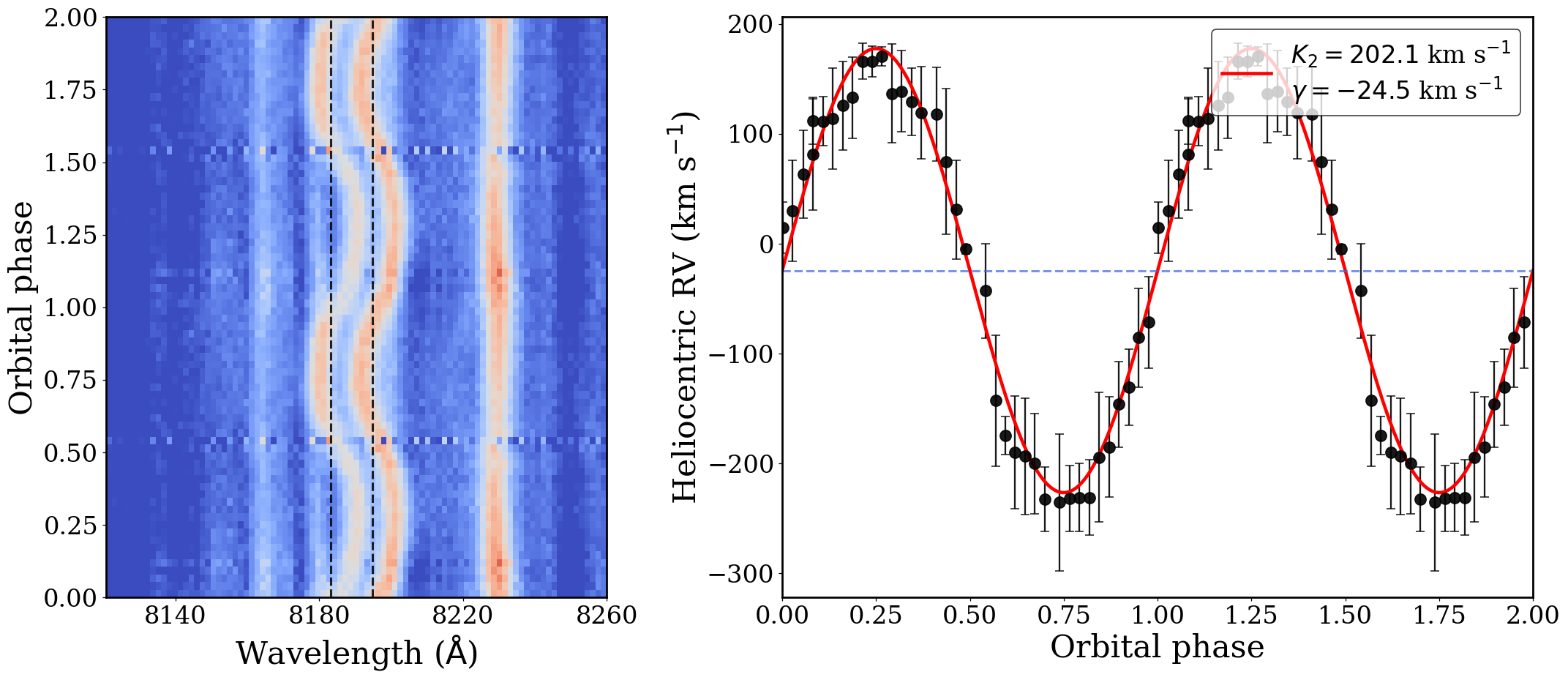}
    \caption{\textit{Left}: Trailed Na~I 8183/8195 doublet in the 38 low-state Palomar/DBSP spectra (9 November 2024), plotted over two orbital cycles. \textit{Right}: Heliocentric donor RVs (\textit{black}) and the median MCMC sinusoidal model (\textit{red}), yielding $K_2 = 202.1\pm 6.5$~km~s$^{-1}$.}
    \label{fig:rv}
\end{figure*}

The trailed Na~I doublet and the RV curve are shown in Figure~\ref{fig:rv}. We fit a sinusoid,
\begin{equation}
    \mathrm{RV}_{\mathrm{donor}} = K_2 \sin\left(2\pi \left(\frac{t - t_0}{P_{\mathrm{orb}}} - \phi_0\right)\right) + \gamma,
    \label{eq:rv}
\end{equation}
with a tight Gaussian prior on $P_{\mathrm{orb}}$ centred on the TESS ephemeris and a per-point jitter, sampled with \texttt{emcee} \citep{2019emcee}. The fit yields
\begin{equation*}
K_2 = 202.1 \pm 6.5~\mathrm{km~s^{-1}}, \qquad \gamma  = -24.5 \pm 4.7~\mathrm{km~s^{-1}},
\end{equation*}
implying a mass function
\begin{equation}
    f(M_{\mathrm{WD}}) = \frac{P_{\mathrm{orb}} K_2^3}{2\pi G} = 0.236^{+0.024}_{-0.021}~M_\odot,
    \label{eq:mass-fn}
\end{equation}
itself a hard lower limit on $M_{\mathrm{WD}}$.

\subsection{Spectral Energy Distribution}
\label{sec:sed}

\begin{figure}
    \centering
    \includegraphics[width=\linewidth]{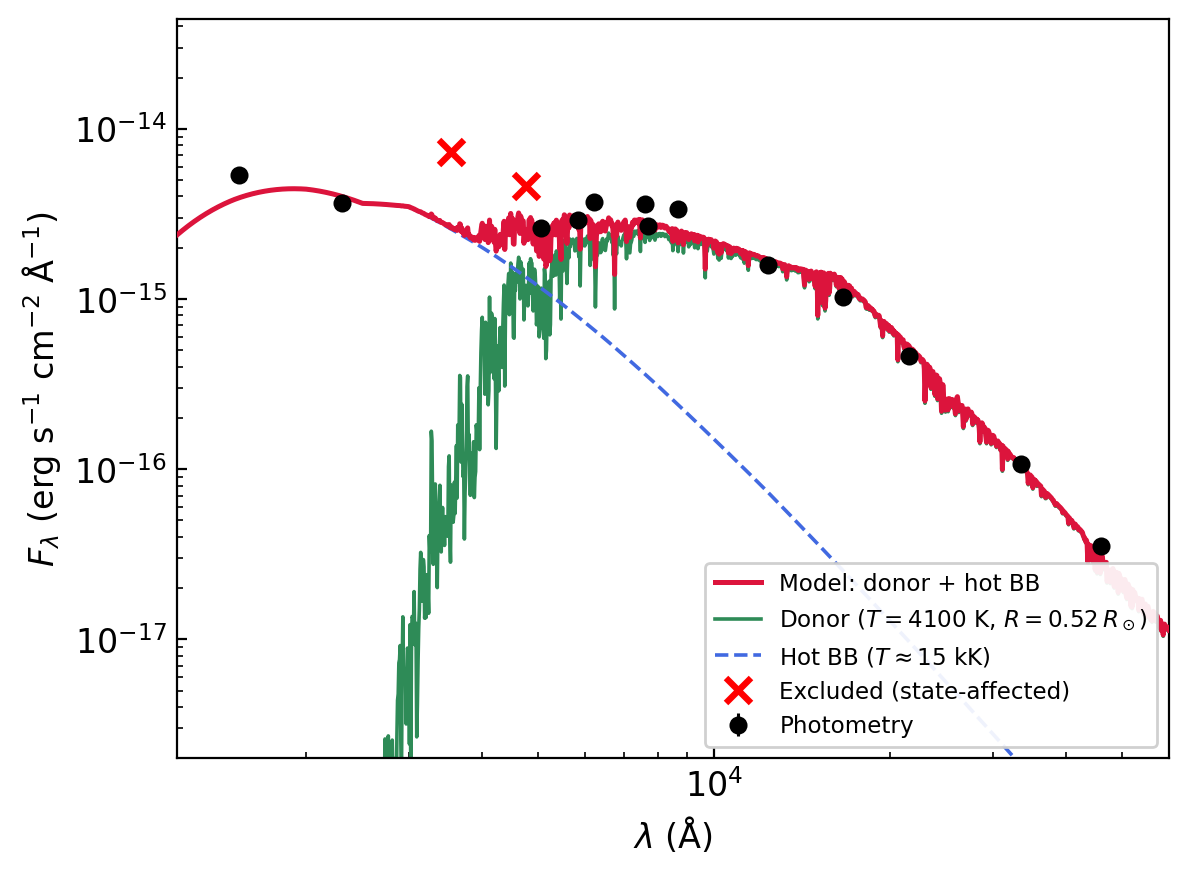}
    \caption{Multi-band SED of Valeria from GALEX FUV/NUV through WISE W2 (\textit{black}; SkyMapper $u/g$ excluded as state-affected, red crosses). The model (\textit{red}) is the adopted donor ($T_{\mathrm{donor}}=4100$~K, $R_{\mathrm{donor}}=0.52~R_\odot$; \textit{green}) plus a hot blackbody fitted to the UV photometry (\textit{blue dashed}, an upper limit given possible non-photospheric UV contributions). The excess of the blue optical bands above the model reflects state-averaged accretion light (Section~\ref{sec:sed}).}
    \label{fig:sed}
\end{figure}

We assemble a broad-band SED from GALEX FUV/NUV, SkyMapper, \textit{Gaia} DR3, 2MASS, and AllWISE photometry, with the distance pinned by the Gaia DR3 parallax ($\varpi = 3.7688 \pm 0.0245$~mas; $d = 265.4 \pm 1.7$~pc). The SkyMapper $u$ and $g$ points sit 5--7$\sigma$ above any donor + WD model; we attribute this to high-state contamination during those visits (in line with the ZTF state changes) and exclude them from the fit. We fit the remaining 13 bands with a BT-Settl/PHOENIX donor \citep{2011allard} plus a hot blackbody, with reddening a free parameter (SFD-anchored Gaussian prior), sampled with \texttt{emcee}. The marginalised posteriors are
\begin{equation*}
\begin{aligned}
T_{\mathrm{donor}} &= 4345 \pm 35~\mathrm{K}, \qquad R_{\mathrm{donor}} = 0.480 \pm 0.009~R_\odot, \\
E(B-V)             &= 0.058 \pm 0.029,
\end{aligned}
\end{equation*}
with the donor parameters anchored by the well-calibrated Gaia, 2MASS, and WISE photometry. The hot component ($T_{\mathrm{WD}} \approx 19{,}000$~K, $R_{\mathrm{WD}}\approx 0.016~R_\odot$) is set by the GALEX detections; since residual UV emission from a propeller-state accretion column can contribute, we treat it as an upper limit ($T_{\mathrm{WD}} \lesssim 22{,}000$~K at 95\%). Figure~\ref{fig:sed} shows the fit.

The SED-only donor temperature must, however, be treated with caution: the survey photometry averages over the high/low state changes (Figure~\ref{fig:ztf}), so blue accretion light contaminates the optical bands -- exactly the effect that forces us to exclude the SkyMapper $u/g$ points -- and biases the fitted donor temperature hot. We therefore checked the temperature directly against the low-state DBSP spectrum, which samples the donor with minimal accretion light: comparing the median low-state spectrum against BT-Settl templates smoothed to the instrumental resolution favours $T_{\mathrm{donor}} = 4100 \pm 150$~K, with the 2MASS colours ($J-K_s = 0.82$) in independent agreement (Appendix~\ref{sec:appendix_donor}).

Because the state-mixing bias is one-sided (blue accretion light can only make the photometry look hotter), we adopt the spectroscopic temperature throughout, and rescale the donor radius to the donor-dominated red/near-IR photometry (SkyMapper $i/z$, Gaia $G_{RP}$, 2MASS, WISE) at the parallax distance:
\begin{equation*}
T_{\mathrm{donor}} = 4100 \pm 150~\mathrm{K}, \qquad R_{\mathrm{donor}} = 0.52 \pm 0.03~R_\odot.
\end{equation*}
As a consistency check, a joint MCMC fit of the SED (with a free jitter term absorbing the state-mixing systematics), the spectroscopic temperature, the RV mass function, and the Roche-filling condition returns intermediate values ($T_{\mathrm{donor}} = 4280^{+70}_{-90}$~K, $R_{\mathrm{donor}} = 0.50\pm0.02~R_\odot$, with a $\approx 15$\% photometric jitter), within $\sim$1$\sigma$ of the adopted solution. Figure~\ref{fig:sed} shows the adopted donor plus a hot blackbody refitted to the UV photometry; the excess of the blue optical bands above this model is the state-mixing contamination.

\subsection{Lower limit on the WD mass from the spin period}
\label{sec:breakup}

\begin{figure}
    \centering
    \includegraphics[width=\linewidth]{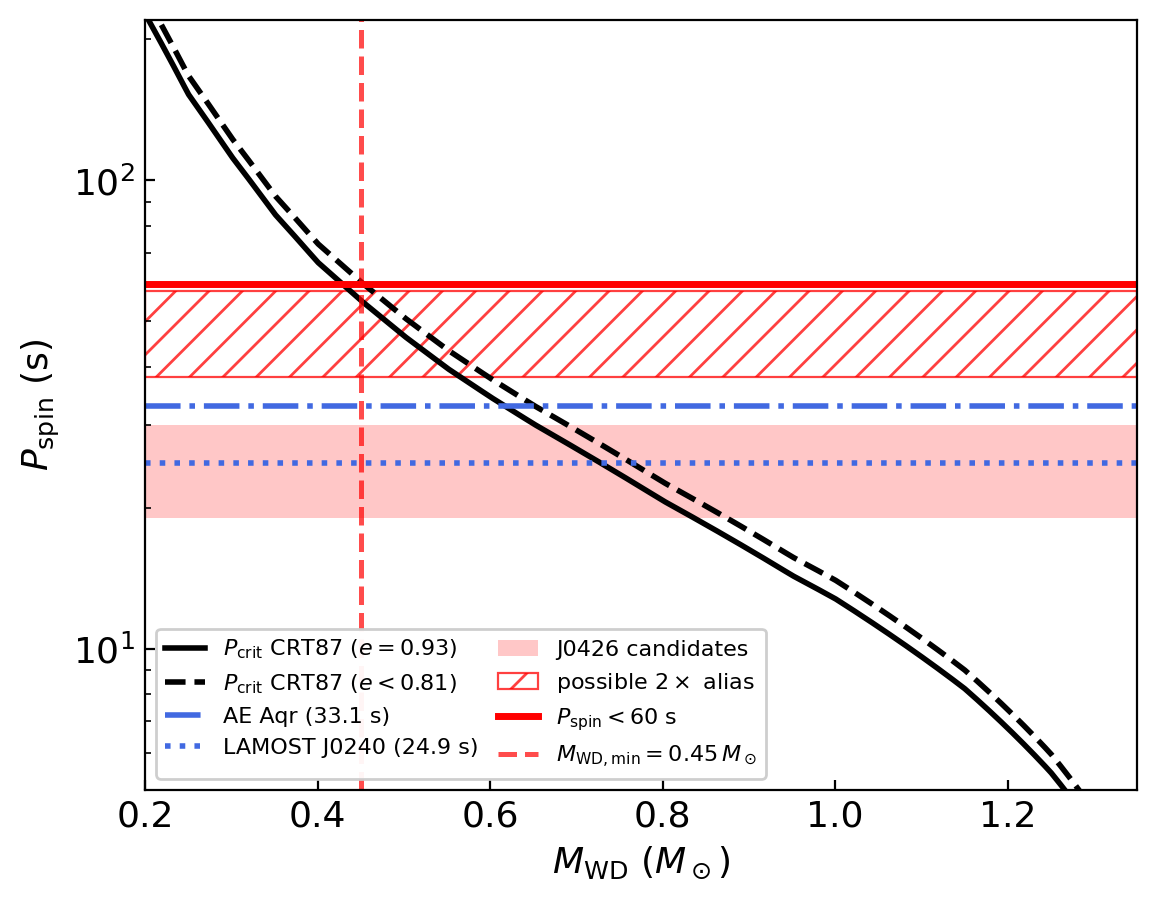}
    \caption{WD spin period vs.\ mass. Black curves show the break-up period (Eq.~\ref{eq:spinmass}) for the \citet{1987chandra_rotating} critical eccentricities, using the \citet{2020bedard} mass--radius relation. The CHIMERA candidate-period band (19--30~s, \textit{solid red shading}), its possible $2\times$ alias band (\textit{hatched}), and the robust limit $P_{\mathrm{spin}}<60$~s (\textit{thick red line}) are shown along with AE~Aqr and LAMOST~J0240. The robust limit implies $M_{\mathrm{WD}}\geq 0.45~M_\odot$; the 19--30~s candidates would imply $M_{\mathrm{WD}}\geq 0.69$--$0.88~M_\odot$.}
    \label{fig:spin_mass}
\end{figure}

Following the analysis presented for LAMOST~J0240 \citep{2025pelisoli_j0240}, the spin period sets an absolute lower limit on the WD mass: a rotating WD becomes unstable to mass loss when its spin frequency exceeds the Keplerian frequency at its equator. Using the rotating-figures result of \citet{1987chandra_rotating}, the maximum spin period at fixed mass is
\begin{equation}
    P_{\mathrm{max}}(M_{\mathrm{WD}}) = \sqrt{\frac{C\,R_{\mathrm{WD}}^3(M_{\mathrm{WD}})}{G M_{\mathrm{WD}}}},
    \label{eq:spinmass}
\end{equation}
with $C \simeq 117$ for the most-flattened (Maclaurin, $e=0.93$) configuration and $C \simeq 141$ for $e<0.81$, combined with the \citet{2020bedard} mass--radius relation. The robust constraint $P_{\mathrm{spin}} < 60$~s then gives
\begin{equation}
    M_{\mathrm{WD}} \geq 0.45~M_\odot,
\end{equation}
while the individual candidate periods imply stronger limits if confirmed: $M_{\mathrm{WD}} \geq 0.69$--$0.88~M_\odot$ for the 19--30~s candidates. Figure~\ref{fig:spin_mass} shows the break-up curves, the candidate-period bands, and the other two WD propellers.

\subsection{No accretion disk: trailed H$\alpha$ and Doppler tomogram}
\label{sec:halpha}

\begin{figure}
    \centering
    \includegraphics[width=\linewidth]{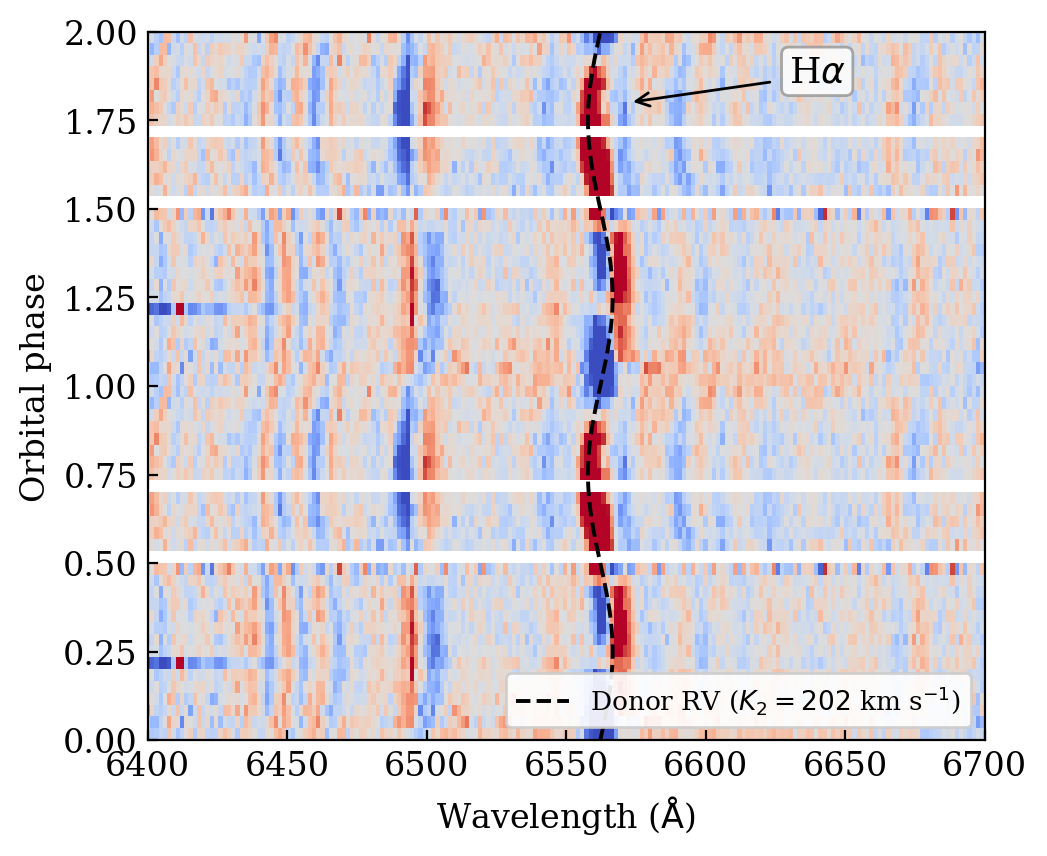}
    \caption{Residual (spectrum $-$ median) trailed spectrum near H$\alpha$ across two orbital cycles. Narrow H$\alpha$ emission traces the donor RV curve (\textit{dashed}), with no broad accretion-disk wings; adjacent photospheric features track the same sinusoid.}
    \label{fig:halpha}
\end{figure}

The low-state spectroscopy confirms the absence of a quiescent accretion disk. Figure~\ref{fig:halpha} shows the residual trailed spectrum (each spectrum minus the orbital median) around H$\alpha$. The line appears in net emission with a narrow ($<200$~km~s$^{-1}$) profile that traces the donor RV curve, with no extended emission at $\pm 1000$--$2000$~km~s$^{-1}$ where a disk's double-peaked profile would lie; neighbouring photospheric Fe~I, Ca~I, and Ti~I features track the same sinusoid. We attribute the narrow H$\alpha$ emission to chromospheric activity on the tidally-locked donor, as seen in many magnetic-CV donors \citep[e.g.][]{1998delfosse, 2017skinner}.

\begin{figure}
    \centering
    \includegraphics[width=\linewidth]{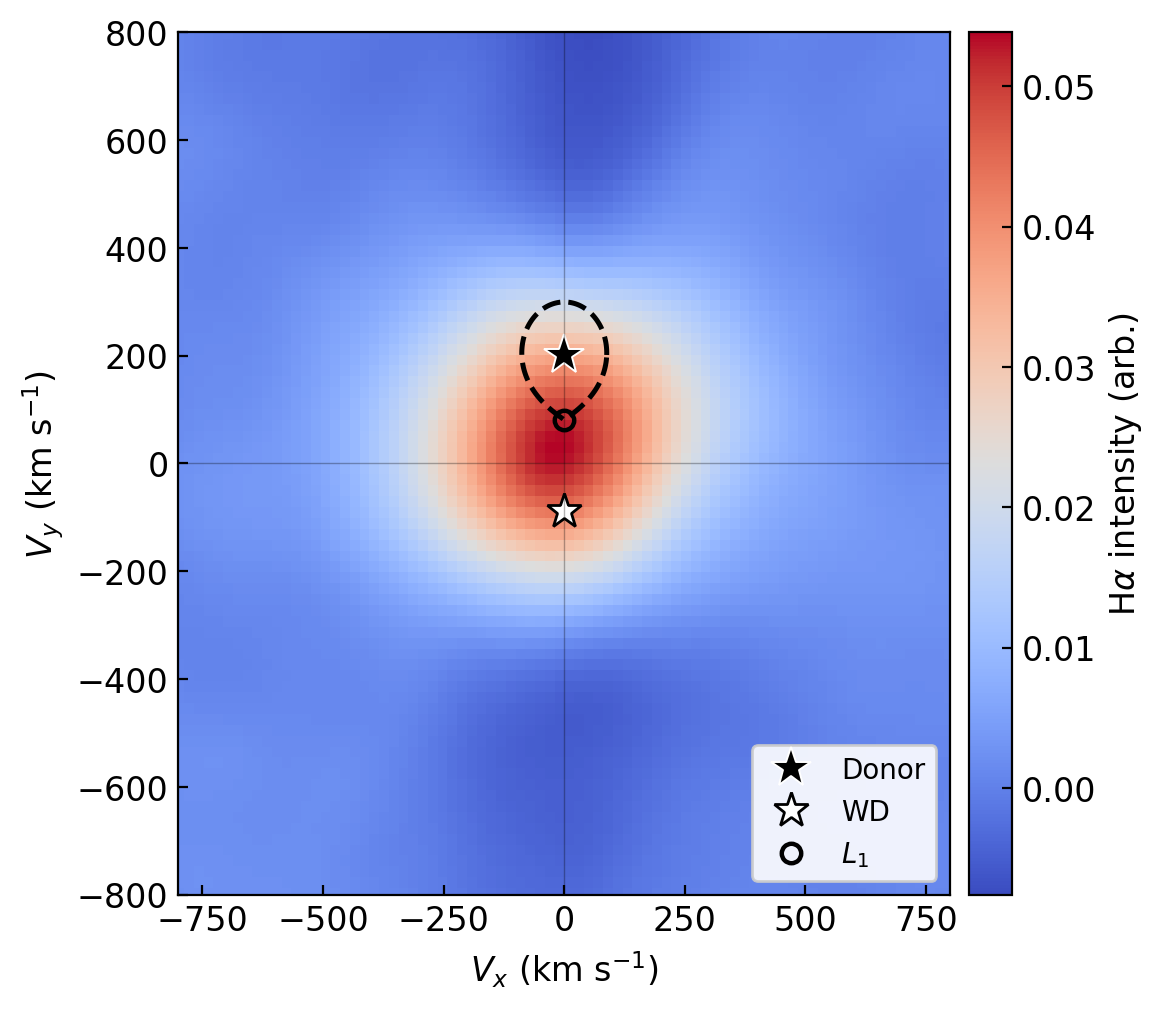}
    \caption{Filtered back-projection Doppler tomogram of the H$\alpha$ emission. The donor Roche lobe (\textit{black dashed contour}) is drawn for fiducial parameters $q=0.45$, $M_{\mathrm{WD}}=0.55~M_\odot$ (its shape varies little over the allowed range); the donor sits at $(0, +K_2)$ (\textit{red star}), the WD at $(0, -K_1)$ (\textit{blue star}), and $L_1$ between them (\textit{white circle}). Emission is concentrated at the donor with the peak offset toward $L_1$. No Keplerian-disk ring (which would appear at $|v - v_{\mathrm{WD}}| \sim 1000$--$4000$~km~s$^{-1}$) and no ballistic stream are detected.}
    \label{fig:tomogram}
\end{figure}

To make this quantitative we compute a Doppler tomogram of H$\alpha$ via filtered back-projection \citep{1988marsh, 2001marsh}, using the 37 good low-state spectra continuum-subtracted over 6480--6650~\AA{} and corrected to the centre-of-mass frame ($\gamma = -24.5$~km~s$^{-1}$). Figure~\ref{fig:tomogram} shows the result: the emission is concentrated in a single compact region away from the donor's velocity-space position $(0, +K_2)$, with the peak offset toward $L_1$ and the centre of mass, consistent with enhanced chromospheric emission on the irradiated, $L_1$-facing hemisphere. Crucially, there is \emph{no} ring of emission centred on the WD velocity -- the signature of a Keplerian disk -- and no curved band tracing an $L_1\to$~WD ballistic stream. Any matter crossing $L_1$ is expelled before it can settle into a disk.

\subsection{Joint binary parameters}
\label{sec:joint}

The following are measured nearly independently of any structural model:
\begin{itemize}
    \item Orbital period: $P_{\mathrm{orb}} = 6.61612 \pm 0.00010$~hr (TESS; Section~\ref{sec:archival}).
    \item Donor RV semi-amplitude: $K_2 = 202.1 \pm 6.5$~km~s$^{-1}$ (Section~\ref{sec:rv}).
    \item WD spin: $P_{\mathrm{spin}} < 60$~s, with candidate detections at 19--30~s (Section~\ref{sec:chimera}).
    \item Distance: $d = 265.4 \pm 1.7$~pc (\textit{Gaia} DR3).
    \item Donor temperature and radius: $T_{\mathrm{donor}} = 4100\pm150$~K (low-state spectrum; Appendix~\ref{sec:appendix_donor}), $R_{\mathrm{donor}} = 0.52\pm0.03~R_\odot$ (red/near-IR photometry + parallax; Section~\ref{sec:sed}).
    \item WD magnetic field: $B_{\mathrm{WD}} = 26\pm3$~MG (Section~\ref{sec:zeeman}).
\end{itemize}

The ongoing mass transfer -- evidenced by the flaring, state changes, and high-state emission lines -- requires the donor to fill its Roche lobe, whose mean density is set purely by the orbital period \citep[e.g.][]{2002frankkingraine}:
\begin{equation}
    \bar\rho_{\mathrm{donor}} \approx 0.185\;\mathrm{g\,cm^{-3}}\,(P_{\mathrm{orb}}/\mathrm{day})^{-2} = 2.46~\mathrm{g\,cm^{-3}}.
    \label{eq:rho-RL}
\end{equation}
Combined with the measured radius, this gives
\begin{equation}
    M_{\mathrm{donor}} = \tfrac{4}{3}\pi R_{\mathrm{donor}}^{3}\,\bar\rho_{\mathrm{donor}} = 0.24 \pm 0.04~M_\odot.
\end{equation}
We regard this as the most reliable donor-mass estimate, because it relies only on the Roche geometry and the measured radius and temperature -- the two donor properties we determine most robustly -- and makes \emph{no} assumption that the donor is a main-sequence star. Indeed, the donor cannot be one: a $0.24~M_\odot$ main-sequence star would have $R \approx 0.26~R_\odot$ and $T \approx 3300$~K, whereas this donor is inflated by a factor of $\sim$2 and $\sim$800~K hotter. A ZAMS star matching the donor temperature (a $\sim 0.7~M_\odot$ K dwarf) is likewise excluded, since its $0.68~R_\odot$ radius is incompatible with the measured $0.52\pm0.03~R_\odot$. The donor has evidently been stripped of much of its mass by past mass transfer onto the WD and is far from thermal equilibrium (potentially still inflating or contracting) as an evolved subgiant; we return to the evolutionary implications in Section~\ref{sec:comparison}. Table~\ref{tab:donormass} compares the adopted solution with the two main-sequence-based alternatives.

\begin{table}[ht]
\caption{Donor-mass estimates and the implied $M_{\mathrm{WD}}$ (for $\sin i = 0.95$). The Roche-fill + measured-radius solution (first row) is adopted; the alternatives assume main-sequence donor structures inconsistent with the measured radius.}
\label{tab:donormass}
\centering
\small
\begin{tabular}{lccc}
\hline\hline
Donor assumption  & $M_{\mathrm{donor}}$ & $R_{\mathrm{donor}}$ & $M_{\mathrm{WD}}$ \\
                  & ($M_\odot$) & ($R_\odot$)  & ($M_\odot$) \\
\hline
Roche-fill + measured $R$ (adopted) & 0.24  & 0.52  & $\sim 0.56$ \\
\hline
\citet{2011knigge} CV donor           & 0.39  & 0.59        & $\sim 0.65$ \\
ZAMS K4V isochrone (MIST)             & 0.70  & 0.68        & $\sim 0.84$ \\
\hline
\end{tabular}
\end{table}

With $M_{\mathrm{donor}} = 0.24~M_\odot$, the binary mass-function relation
\begin{equation}
    K_2 = \left(\frac{2\pi G}{P_{\mathrm{orb}}}\right)^{1/3}\!\frac{M_{\mathrm{WD}} \sin i}{(M_{\mathrm{WD}}+M_{\mathrm{donor}})^{2/3}}
    \label{eq:k2}
\end{equation}
maps inclination onto WD mass: $M_{\mathrm{WD}} = 0.51~M_\odot$ at $i=90^{\circ}$, $0.56~M_\odot$ at $72^{\circ}$, $0.67~M_\odot$ at $60^{\circ}$, and $0.86~M_\odot$ at $50^{\circ}$; marginalising over a uniform $\cos i$ inclination distribution (propagating the $K_2$ and $R_{\mathrm{donor}}$ uncertainties, and requiring $M_{\mathrm{WD}} < 1.44~M_\odot$) gives $M_{\mathrm{WD}} = 0.61^{+0.31}_{-0.10}~M_\odot$. Eclipses provide no useful constraint on the inclination: the WD contributes only a small fraction of the optical light (Section~\ref{sec:sed}), so even an edge-on system would show at most a shallow eclipse, undetectable amid the flaring. The large amplitude of the $P_{\mathrm{orb}}/2$ modulation in TESS does, however, qualitatively favour a high inclination, since both the photospheric ellipsoidal component and any orbitally-modulated accretion light scale as $\sin^2 i$ (Appendix~\ref{sec:appendix_tess}) -- pushing $M_{\mathrm{WD}}$ toward the lower end of the range, $\approx 0.55~M_\odot$. The entire range is consistent with the break-up limit $M_{\mathrm{WD}} \geq 0.45~M_\odot$ from $P_{\mathrm{spin}}<60$~s. If the 19--30~s candidate spin periods are confirmed, the stronger break-up limit ($M_{\mathrm{WD}} \geq 0.69$--$0.88~M_\odot$) would require $i \lesssim 49^{\circ}$--$58^{\circ}$ -- fully consistent with the joint-fit posterior, though in tension with the qualitative high-inclination preference; a definitive spin measurement would resolve this. The corresponding Roche-lobe radius \citep{1983eggleton} is $R_L = 0.51\pm0.03~R_\odot$ across the posterior -- in excellent agreement with the independently measured radius of $0.52\pm0.03~R_\odot$, a strong consistency check on the adopted solution.

Our adopted binary parameters are then
\begin{equation*}
\begin{aligned}
M_{\mathrm{donor}}    &= 0.24 \pm 0.04~M_\odot,\\
R_{\mathrm{donor}}    &= R_L = 0.52 \pm 0.03~R_\odot,\\
M_{\mathrm{WD}}       &= 0.61^{+0.31}_{-0.10}~M_\odot,\\
q = M_{\mathrm{donor}}/M_{\mathrm{WD}} &= 0.39\pm0.11,\\
a                     &\approx 1.7~R_\odot,
\end{aligned}
\end{equation*}
summarised with their provenance in Table~\ref{tab:params}.

\begin{table*}
\caption{Inferred parameters of 1eRASS~J042633.5$-$250226. Adopted binary parameters use the spectroscopic donor temperature, the red/near-IR radius, and the Roche-filling condition (\S\ref{sec:sed}, \S\ref{sec:joint}). \label{tab:params}}
\centering
\small
\begin{tabular}{lc l}
\hline\hline
Parameter & Value & Source \\
\hline
\multicolumn{3}{l}{\textit{Astrometry and photometry}} \\
Gaia DR3 source ID & 4896134903510805888 & Gaia DR3 \\
RA, Dec (J2000)    & 04:26:33.66, $-25^{\circ}02'28.5''$ & Gaia DR3 \\
$\varpi$ (mas)     & $3.7688 \pm 0.0245$ & Gaia DR3 \\
$d$ (pc)           & $265.4 \pm 1.7$ & Gaia parallax \\
$G$ (mag)          & $14.98$ & Gaia DR3 \\
$E(B-V)$           & $0.058 \pm 0.029$ & SED \\
$F_X \textrm{ (erg s$^{-1}$ cm$^{-2}$)}$ & $8.61\times10^{-13}$ & SRG/eROSITA (0.2--2.3 keV)\\
$L_X \textrm{ (erg s$^{-1}$)}$ & $6.84\times10^{30}$ & SRG/eROSITA (0.2--2.3 keV)\\
$F_\textrm{radio}\;(\mu\textrm{Jy})$ & $< 13.4$ & VLA C-band (3$\sigma$ upp. limit)\\
\hline
\multicolumn{3}{l}{\textit{Ephemeris}} \\
$P_{\mathrm{orb}}$ (hr)          & $6.61612 \pm 0.00010$ & TESS (\S\ref{sec:archival}) \\
$T_0$ (BJD$_{\mathrm{TDB}}$; photometric minimum) & $2459144.6375 \pm 0.0002$ & TESS \\
$P_{\mathrm{spin}}$ (s)          & $<60$; candidates 19--30 (strongest 26.8) & CHIMERA (\S\ref{sec:chimera}) \\
\hline
\multicolumn{3}{l}{\textit{Radial-velocity solution}} \\
$K_2$ (km s$^{-1}$)              & $202.1 \pm 6.5$ & RV (\S\ref{sec:rv}) \\
$\gamma$ (km s$^{-1}$)           & $-24.5 \pm 4.7$ & RV \\
$f(M_{\mathrm{WD}})$ ($M_\odot$) & $0.236^{+0.024}_{-0.021}$ & RV \\
\hline
\multicolumn{3}{l}{\textit{Adopted stellar and binary parameters}} \\
$T_{\mathrm{donor}}$ (K)         & $4100 \pm 150$ & low-state spectrum (App.~\ref{sec:appendix_donor}) \\
$R_{\mathrm{donor}}$ ($R_\odot$) & $0.52 \pm 0.03$ & red/near-IR SED + parallax (\S\ref{sec:sed}) \\
$M_{\mathrm{donor}}$ ($M_\odot$) & $0.24 \pm 0.04$ & Roche condition (Eq.~\ref{eq:rho-RL}) \\
$M_{\mathrm{WD}}$ ($M_\odot$)    & $0.61^{+0.31}_{-0.10}$ (joint constraint); $\geq 0.45$ (break-up) & \S\ref{sec:joint}; \S\ref{sec:breakup} \\
$i$ (deg)                        & unconstrained; high $i$ favoured by orbital modulation & \S\ref{sec:joint}; App.~\ref{sec:appendix_tess} \\
$q = M_{\mathrm{donor}}/M_{\mathrm{WD}}$ & $0.39 \pm 0.11$ & \S\ref{sec:joint} \\
$a$ ($R_\odot$)                  & $1.70^{+0.18}_{-0.09}$ & Kepler's law \\
$R_L$ ($R_\odot$)                & $0.51 \pm 0.03$ (consistent with $R_{\mathrm{donor}}$) & \citet{1983eggleton} \\
$T_{\mathrm{WD}}$ (K)            & $\lesssim 22{,}000$ (upper limit) & SED \\
$B_{\mathrm{WD}}$ (MG)           & $26 \pm 3$ & Zeeman (\S\ref{sec:zeeman}) \\
\hline
\end{tabular}
\end{table*}

\subsection{Magnetic field from Zeeman splitting}
\label{sec:zeeman}

The low-state spectrum -- in which the accretion light is at its weakest -- shows several narrow absorption features flanking the positions of the Balmer lines that are not obviously attributable to the donor photosphere, which we interpret as Zeeman-split components of the magnetised WD. At fields of tens of MG the linear Zeeman picture is inadequate -- quadratic Zeeman and Lorentz-mixing effects produce many additional transitions -- so we use the full hydrogen Zeeman-pattern wavelengths of \citet{2019zhao_liu}.

\begin{figure*}
    \centering
    \includegraphics[width=0.95\linewidth]{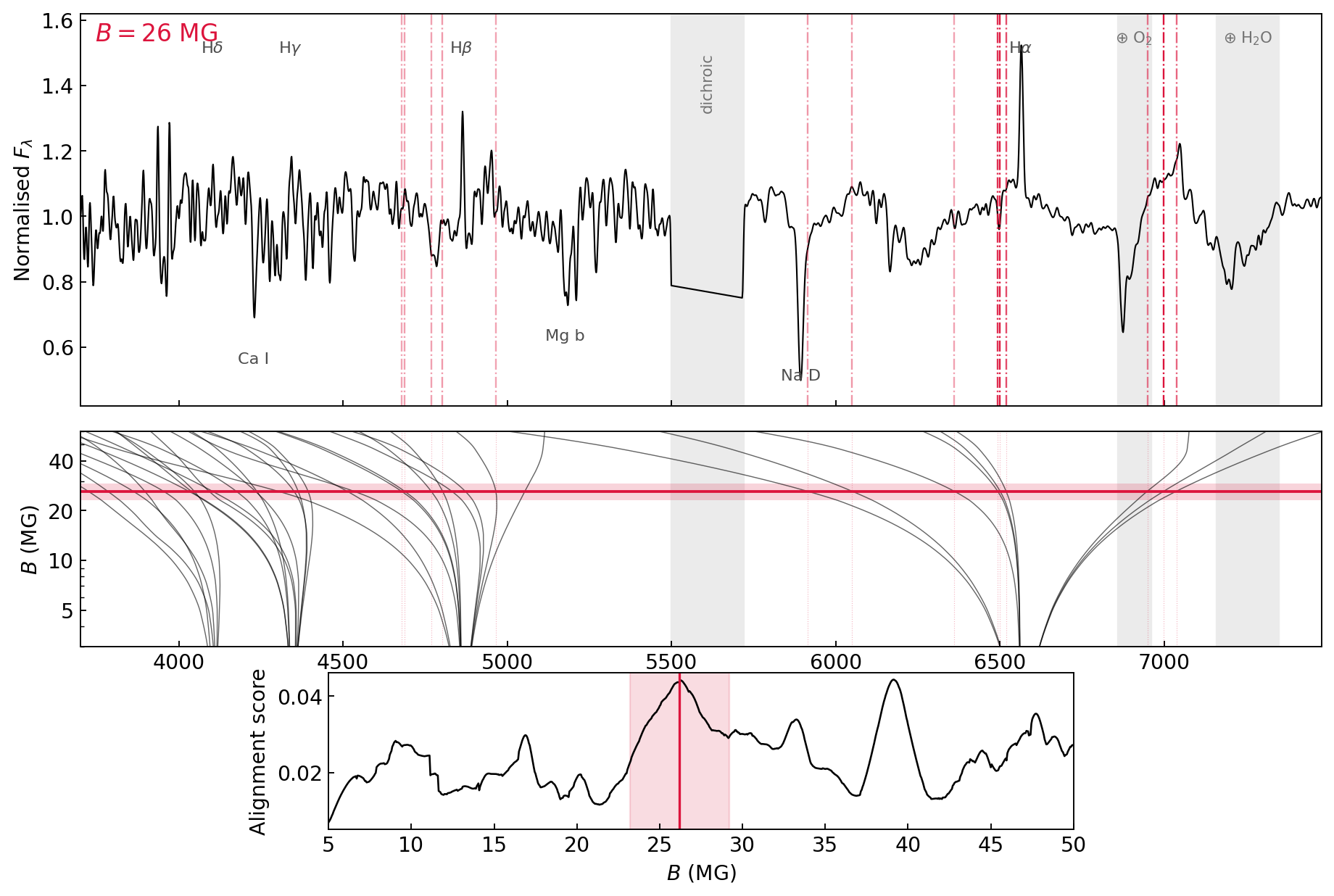}
    \caption{Zeeman analysis of the average low-state DBSP spectrum. \textit{Top}: median continuum-normalised low-state spectrum (\textit{black}) with the hydrogen Zeeman components predicted at the adopted $B = 26$~MG by \citet{2019zhao_liu} (\textit{red dash-dotted lines}, opacity scaled by oscillator strength); the main donor photospheric features (Ca~I, Mg~b, Na~D), the telluric O$_2$ and H$_2$O bands ($\oplus$), and the dichroic crossover are labelled. \textit{Middle}: component wavelengths as functions of field strength, with the adopted field marked (\textit{red line and band}). \textit{Bottom}: oscillator-strength-weighted alignment score between the predicted component positions and the detected absorption features as a function of $B$, maximised at $B \approx 26$~MG (adopted $26\pm3$~MG), with a comparably strong secondary maximum at 33~MG (Appendix~\ref{sec:appendix_zeeman33}) and a weaker alias near 40~MG (in the sense that it is dominated by the splitting of just the H$\alpha$ line).}
    \label{fig:zeeman}
\end{figure*}

To determine the field quantitatively, we detect the absorption features in the continuum-normalised median low-state spectrum (masking emission lines, telluric bands, the Na~D complex, and the dichroic crossover) and, for each trial field over $B = 5$--$50$~MG, score the alignment between the detected features and the predicted Balmer component wavelengths, weighting each component by its oscillator strength and each feature by its depth. The alignment score (Figure~\ref{fig:zeeman}, bottom panel) is maximised at $B \approx 26$~MG -- with the strongest coincidences at 4784, 4832, 4922, and 6231~\AA{} -- with a comparably strong secondary maximum at $B \approx 33$~MG and a weaker alias near 40~MG. The same analysis applied to the high-state spectrum yields a consistent optimum near $30$~MG, anchored by an exact coincidence of the predicted H$\beta$ $\sigma$ component with the strong absorption feature at 4782~\AA{}. We adopt the global low-state optimum,
\begin{equation*}
B_{\mathrm{WD}} = 26 \pm 3~\mathrm{MG},
\end{equation*}
and present the 33~MG solution as an alternative in Appendix~\ref{sec:appendix_zeeman33}. 

The uncertainty reflects the width of the alignment peak, the interpolation of the \citet{2019zhao_liu} tables (which sample the field coarsely between 23.5 and 47~MG), and the unknown viewing geometry and field structure. We caution that the dominant systematic is the identification of the features themselves: many low-state absorption features arise in the donor photosphere, and the alignment statistic alone does not single out a unique field at high formal significance -- the value above rests on the attribution of the flanking features to the WD, with the 33~MG maximum (and, more weakly, the $\sim$40~MG alias) as the principal alternatives. Figure~\ref{fig:zeeman} shows the low-state spectrum with the predicted component positions at the adopted field, and the component wavelengths as functions of $B$. This places the magnetic field strength 1eRASS J0426 solidly in the regime of polars, which have magnetic fields in the range of 10--240 MG. However, we caution that the combination of noise in the spectra along with uncertainty in the detection of the individual Zeeman splits prevents us from placing a confident constraint on the magnetic field of 1eRASS J0426. Future (higher resolution) spectroscopic observations are needed to solidify this measurement.

\subsection{Confirmation of propeller mode}
\label{sec:propellermode}

A magnetised WD is in the propeller (matter-ejecting) regime when the magnetospheric (Alfv\'en) radius
\begin{equation}
    R_A = \left(\frac{\mu^4}{2\,G\,M_{\mathrm{WD}}\,\dot{M}^{\,2}}\right)^{1/7}, \qquad \mu = B_{\mathrm{WD}}R_{\mathrm{WD}}^{3},
\end{equation}
exceeds the corotation radius $R_{\mathrm{co}} = (G\,M_{\mathrm{WD}}\,P_{\mathrm{spin}}^{\,2}/4\pi^{2})^{1/3}$ -- equivalently, when $P_{\mathrm{spin}}$ is shorter than the equilibrium period $P_{\mathrm{eq}} = 2\pi(R_A^{3}/GM_{\mathrm{WD}})^{1/2}$. For a fiducial $M_{\mathrm{WD}} = 0.55~M_\odot$ ($R_{\mathrm{WD}} = 0.0135~R_\odot$; \citealt{2020bedard}), $B = 26$~MG, and $\dot{M} = 5\times 10^{-10}~M_\odot$~yr$^{-1}$ \citep[the][secular rate at this period]{2011knigge}, the magnetospheric radius is $R_A \approx 2.1~R_\odot$ -- formally larger than the binary separation itself -- while $R_{\mathrm{co}} \leq 0.027~R_\odot$ for any $P_{\mathrm{spin}} < 60$~s, giving $R_A/R_{\mathrm{co}} \gtrsim 78$ and $P_{\mathrm{eq}} \approx 4.1\times 10^{4}$~s.

\begin{figure}
    \centering
    \includegraphics[width=\linewidth]{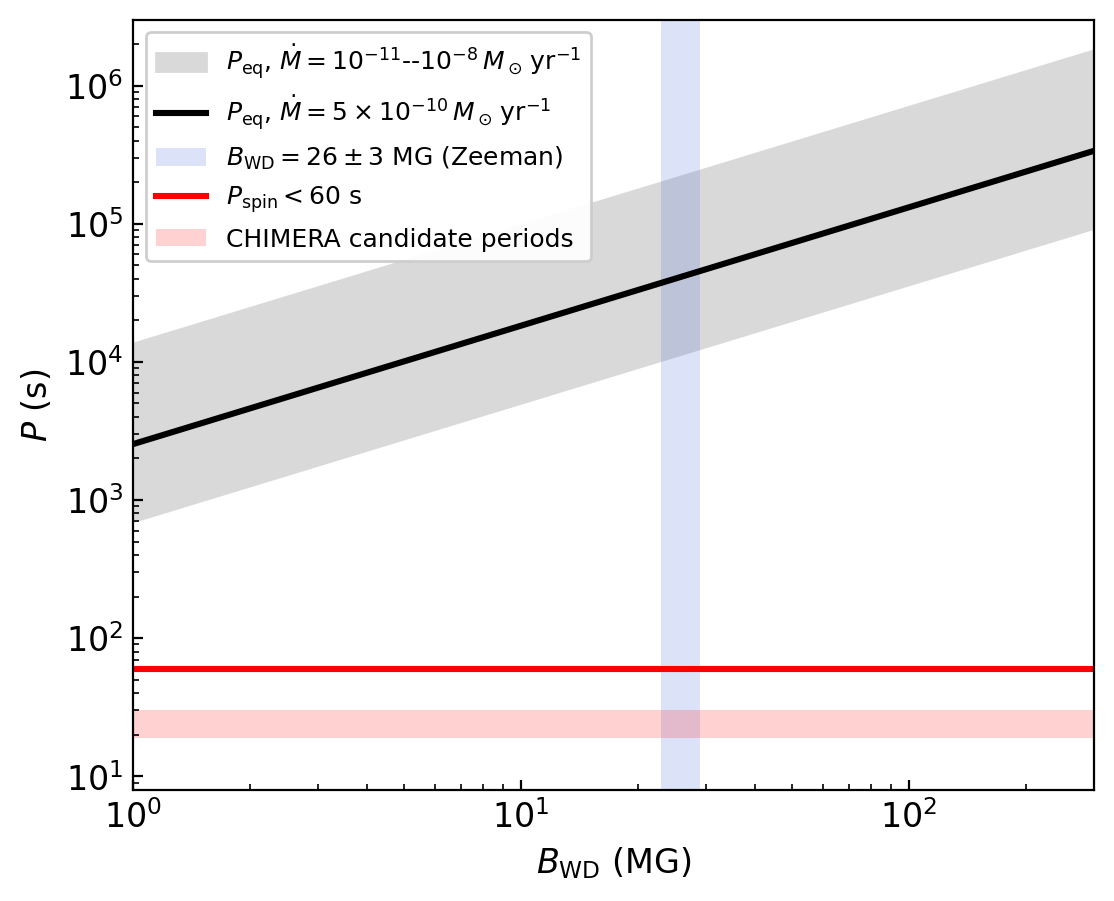}
    \caption{Equilibrium spin period $P_{\mathrm{eq}}$ vs.\ WD magnetic field (for a fiducial $M_{\mathrm{WD}}=0.55~M_\odot$), for the fiducial \citet{2011knigge} mass-transfer rate at $P_{\mathrm{orb}}=6.6$~hr (\textit{black line}) and the wider range $\dot{M} = 10^{-11}\text{--}10^{-8}~M_\odot$~yr$^{-1}$ (\textit{grey band}). At the Zeeman-inferred $B_{\mathrm{WD}} = 26\pm3$~MG (\textit{blue band}), any spin period below the robust 60-s limit (\textit{red line}; candidate periods in \textit{red band}) is orders of magnitude shorter than $P_{\mathrm{eq}}$: the system must be a propeller.}
    \label{fig:propmode}
\end{figure}

Figure~\ref{fig:propmode} shows $P_{\mathrm{eq}}(B,\dot{M})$ for $10^{-11} < \dot{M} < 10^{-8}~M_\odot$~yr$^{-1}$. Even the loosest spin constraint, $P_{\mathrm{spin}} = 60$~s, lies more than two orders of magnitude below $P_{\mathrm{eq}}$ at $B = 26$~MG for all plausible $\dot{M}$; pushing $\dot{M}$ to a polar-like $10^{-7}~M_\odot$~yr$^{-1}$ still leaves more than an order of magnitude. \emph{1eRASS~J0426 is unambiguously in the propeller regime.}

\section{Discussion}
\label{sec:discussion}

\subsection{The propeller population: stripped, out-of-equilibrium donors}
\label{sec:comparison}

The discovery of 1eRASS~J0426 brings the population of magnetic WD propellers to three. All share long binary periods (6.6--9.9~hr), hot K-type Roche-lobe-filling donors, low X-ray luminosities, large-amplitude optical flaring, and ultra-fast WD spins.

\begin{figure}
    \centering
    \includegraphics[width=\linewidth]{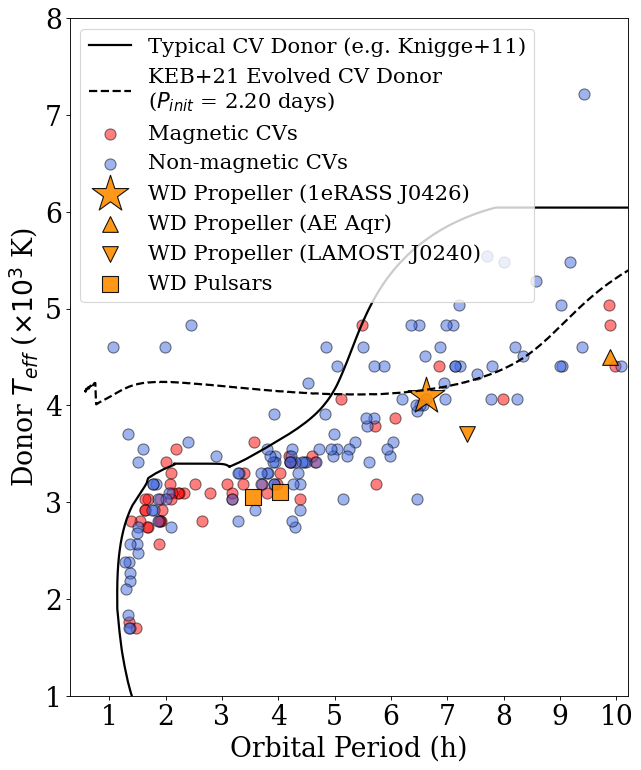}
    \caption{Orbital period vs.\ donor effective temperature for CVs from the 2015 version of the catalog of \cite{2003ritter}. The three known WD propellers (orange star and triangles) cluster at $P_{\mathrm{orb}}\approx 6.6$--$9.9$~hr and $T_{\mathrm{donor}}\approx 4000$--$4500$~K, consistent with stripped, out-of-thermal-equilibrium donors \citep[e.g.][]{2021el-badry}. They likely came into contact at longer $\gtrsim 8$hr orbital periods than ``standard'' CVs described by the tracks of \cite{2011knigge}.}
    \label{fig:porb_tdonor}
\end{figure}

Figure~\ref{fig:porb_tdonor} places the three propellers in the $P_{\mathrm{orb}}$--$T_{\mathrm{donor}}$ plane alongside magnetic and non-magnetic CVs, WD pulsars, and the standard CV donor sequence of \citet{2011knigge}. The propellers cluster tightly at $T_{\mathrm{donor}}=4000$--$4500$~K. This is expected for evolved donors that come into contact at longer orbital periods $P_\textrm{orb}\gtrsim 8$ hr, as shown by binary stellar evolution models in \cite{2021el-badry}. 

The binary solution of Section~\ref{sec:joint} indicates why. The donor of 1eRASS~J0426 is simultaneously \emph{undermassive} ($0.24~M_\odot$, vs.\ $0.39~M_\odot$ on the standard track at this period) and \emph{hot and inflated} for its mass: it resembles the stripped remnant of an initially more massive star rather than a slowly-whittled main-sequence donor. Precisely this configuration is the established interpretation of AE~Aqr, whose donor's anomalous surface abundances and luminosity record a phase of rapid, thermal-timescale mass transfer from an initially more massive companion \citep{2002schenker}; a high-$\dot M$ episode of this kind is also the natural way to spin the WD up to near break-up. In this picture, the WD propellers are systems caught shortly \emph{after} such an episode: the spun-up, magnetised WD now ejects the much-reduced mass-transfer stream, while the donor -- stripped and evolved far from thermal equilibrium -- is still relaxing (inflating or contracting) toward the structure appropriate to its new, lower mass. As the propelled ejecta drain the WD's spin angular momentum, the system should eventually transition to ordinary intermediate-polar accretion and migrate into the bulk IP population.

\subsection{Space density and birth rate}
\label{sec:density}

1eRASS~J0426 ($d=265$~pc) is the second-nearest WD propeller after AE~Aqr (91.6~pc); LAMOST~J0240 lies at 620~pc. Using the $1/V_{\mathrm{max}}$ estimator \citep{1993tinney} with an exponential disk correction corresponding to a scale height of 200 pc \citep[values ranging between 100 pc and 1 kpc do not appreciably change the final calculations; e.g. ][]{2024rodriguez_survey}, the three systems give
\begin{equation*}
\rho_{\mathrm{prop}} \gtrsim 3.4\times 10^{-7}~\mathrm{pc^{-3}},
\end{equation*}
dominated by AE~Aqr, though now with smaller error bars thanks to the three-system population. Since none of the three was found in a complete unbiased survey, this is a robust lower limit. For comparison, the local space density of magnetic CVs is $\rho_{\mathrm{mCV}}\approx 1.5\times 10^{-6}~\mathrm{pc}^{-3}$ \citep[from the volume-limited samples of][]{2020pala, 2024rodriguez_survey}: WD propellers constitute at least $\sim$20\% of the magnetic-CV space density.

\begin{figure}
    \centering
    \includegraphics[width=\linewidth]{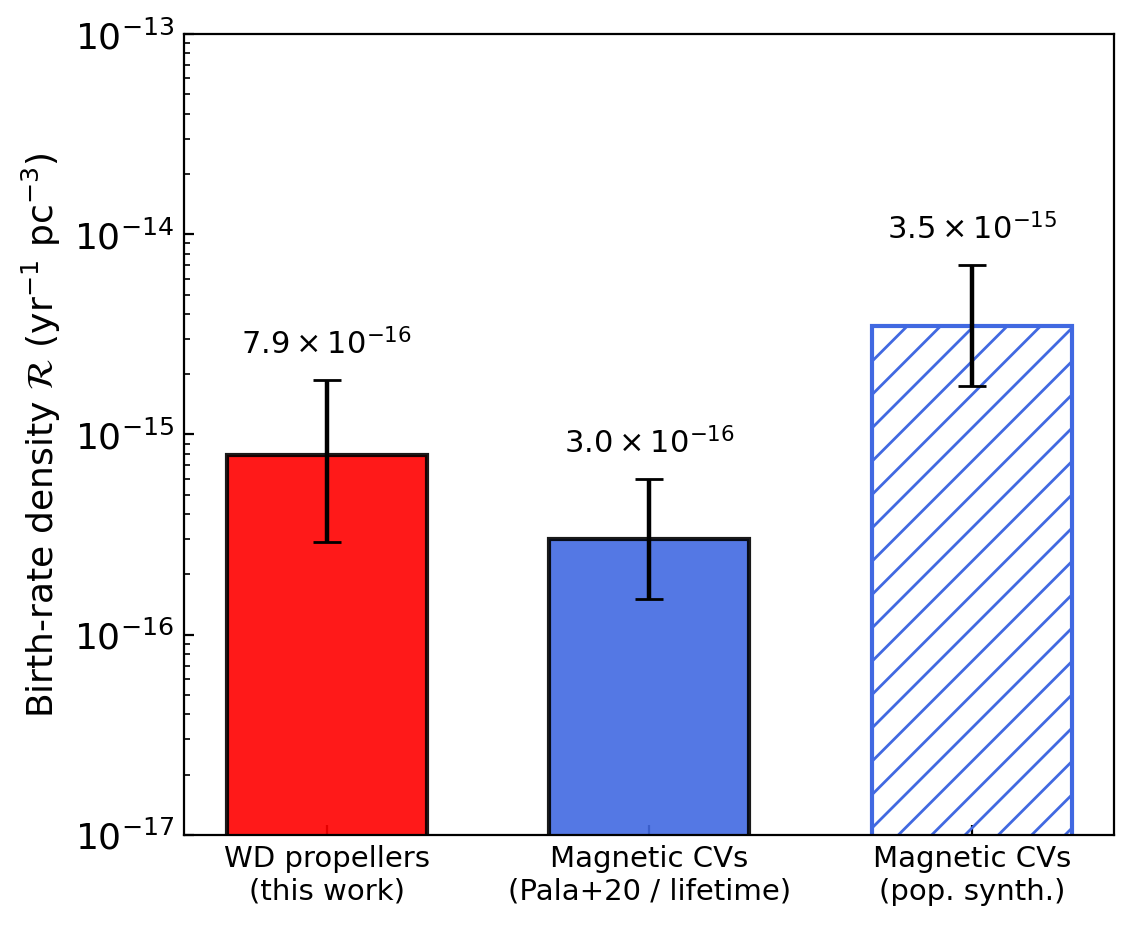}
    \caption{Implied birth-rate densities $\mathcal{R}$. The WD-propeller birth rate (\textit{red}; Eq.~\ref{eq:birthrate}, using $\Delta t_{\mathrm{prop}}\approx 425$~Myr in the 6--10~hr window along the \citealt{2011knigge} track where WD propellers are found) is comparable to the magnetic-CV birth rate from \citet{2020pala} (\textit{solid blue}) and \citet{2015goliasch} (\textit{hatched blue}). Error bars show the dominant uncertainties: Poisson counting statistics of the three known propellers (though the WD propeller estimates should be treated as a lower limit), and factor-of-two lifetime/space-density systematics for the magnetic-CV rates.}
    \label{fig:birthrate}
\end{figure}

Dividing by the time systems spend in the observed orbital-period window converts this into a birth rate. Integrating along the \citet{2011knigge} track from $P_{\mathrm{orb}} = 9.9$~hr to 6.6~hr (the general orbital period range in which WD propellers are found) gives $\Delta t_{\mathrm{prop}} \approx 425$~Myr, so
\begin{equation}
    \mathcal{R}_{\mathrm{prop}} \gtrsim 8\times 10^{-16}~\mathrm{yr}^{-1}~\mathrm{pc}^{-3},
    \label{eq:birthrate}
\end{equation}
again a conservative lower limit, since in the evolved CV tracks presented by \cite{2021el-badry}, the evolution in this orbital period range proceeds even faster. The corresponding magnetic-CV birth rate is $\mathcal{R}_{\mathrm{mCV}}\approx 3\times 10^{-16}$~yr$^{-1}$~pc$^{-3}$, from the \citet{2020pala} space density and a $\sim$5~Gyr lifetime, or $\sim 3.5\times 10^{-15}$~yr$^{-1}$~pc$^{-3}$ from population synthesis \citep[scaling the][CV birth rate by the magnetic fraction]{2015goliasch}. Both straddle the propeller value within a factor of a few (Figure~\ref{fig:birthrate}): \textit{the birth rate of WD propellers is comparable to that of magnetic CVs}, arguing that a substantial fraction -- perhaps most -- magnetic CVs pass through a propeller phase early in their evolution, before sustained accretion spins the WD down toward equilibrium. Consistency with AE~Aqr's rapid spin-down \citep[$\dot P_{\mathrm{spin}}\approx 5.6\times 10^{-14}$~s~s$^{-1}$, implying a $\sim 10^{7}$-yr timescale;][]{1997wynn} requires that the WD's angular momentum be episodically replenished by bursts of mass transfer -- precisely what the high/low state changes seen in ZTF (Figure~\ref{fig:ztf}) and in AE~Aqr \citep{2017buckley} appear to encode.

\subsection{State changes}
\label{sec:states}

The ZTF photometry (Figure~\ref{fig:ztf}) shows month-long high and low states, with the system $\sim$0.5~mag brighter and flaring at up to 30\% amplitude in the high state. The high-state SkyMapper excess (Section~\ref{sec:sed}) leaves the same fingerprint in the UV/blue continuum. Our VLA observation, taken during a low state, did not detect the system; coordinated radio + optical monitoring during a high state would test whether the propelled ejecta produce detectable synchrotron emission, as suggested for AE~Aqr \citep{2015coppejans}.

\section{Conclusion}
\label{sec:conclusion}

We have presented the discovery of 1eRASS~J042633.5$-$250226, the third magnetic WD propeller, at only $265.4 \pm 1.7$~pc. It shares all the hallmarks of the prototype AE~Aqr: orbital modulation and high-amplitude flaring in TESS, month-long state changes in ZTF, low and hard X-ray luminosity, an H$\alpha$ Doppler tomogram free of any disk or stream, and coherent short-period optical pulsations. The pulsations, detected on three CHIMERA nights at candidate periods of 19--30~s (with the strongest being at 26.8 sec; possibly twice that if harmonic-dominated), establish $P_{\mathrm{spin}} < 60$~s -- among the fastest WD spins known -- and a break-up limit of $M_{\mathrm{WD}} \geq 0.45~M_\odot$.

Phase-resolved spectroscopy yields $K_2 = 202.1 \pm 6.5$~km~s$^{-1}$ and $f(M_{\mathrm{WD}}) = 0.236~M_\odot$ at the 6.6161-hr orbital period, allowing for a nearly complete solution of binary parameters. The donor -- hot, undermassive, and inflated -- is an evolved star, as established for AE~Aqr's donor following thermal-timescale mass transfer. Zeeman splitting gives $B_{\mathrm{WD}} = 26\pm3$~MG, for which any spin period below 60~s is orders of magnitude faster than equilibrium: the system is unambiguously a propeller. This also marks the first time that the magnetic field of a WD propeller has been directly measured, confirming the connection between such systems and polars.

The proximity of 1eRASS~J0426 implies a local space density of WD propellers of $\rho \gtrsim 3\times 10^{-7}~\mathrm{pc}^{-3}$ and a birth rate comparable to that of magnetic CVs, suggesting the propeller phase is a common early stage of magnetic-CV evolution. Pinning down the WD spin period with a dedicated multi-night high-speed campaign, dedicated X-ray observations to detect the spin at keV energies, and catching the system in a radio-loud high state, are the clear next steps. 

The discovery of 1eRASS~J0426 solidifies WD propellers as a population of systems with ultra-rapidly spinning magnetized WDs. Their long orbital periods, at the highest end of CV orbital periods (Figure \ref{fig:donor_temp}), signify that they are young systems, with Roche lobe overflow having started recently. Intriguingly, this could be strong evidence that WD magnetism emerges as a result of rapid, accretion-induced rotation. The continued, detailed study of such systems and discovery of new ones will be imperative to test the evolutionary story of WD magnetic field formation. With future SRG/eROSITA releases on the horizon, continued optical photometric monitoring by ZTF and TESS, and now the Rubin Legacy Survey of Space and Time (LSST), more such systems are expected be discovered in years to come.

\section*{Acknowledgments}
ACR acknowledges support from a Future Faculty Leader Postdoctoral Fellowship and the Institute for Theory and Computation at the Center for Astrophysics $|$ Harvard \& Smithsonian. We thank the staffs of Palomar and Lick Observatories for their assistance in acquiring data presented here. We are also grateful to the VLA DDT committee for approving discretionary time on a rapid basis. This work was supported by NSF grant AST-2540180.

This work is based on data from eROSITA, the soft X-ray instrument aboard SRG, a joint Russian-German science mission supported by the Russian Space Agency (Roskosmos), in the interests of the Russian Academy of Sciences represented by its Space Research Institute (IKI), and the Deutsches Zentrum f{\"u}r Luft- und Raumfahrt (DLR). The SRG spacecraft was built by Lavochkin Association (NPOL) and its subcontractors, and is operated by NPOL with support from the Max Planck Institute for Extraterrestrial Physics (MPE). The development and construction of the eROSITA X-ray instrument was led by MPE, with contributions from the Dr. Karl Remeis Observatory Bamberg \& ECAP (FAU Erlangen-Nuernberg), the University of Hamburg Observatory, the Leibniz Institute for Astrophysics Potsdam (AIP), and the Institute for Astronomy and Astrophysics of the University of T{\"u}bingen, with the support of DLR and the Max Planck Society. The Argelander Institute for Astronomy of the University of Bonn and the Ludwig Maximilians Universit{\"a}t Munich also participated in the science preparation for eROSITA.

This work presents results from the European Space Agency (ESA) space mission Gaia. Gaia data are being processed by the Gaia Data Processing and Analysis Consortium (DPAC). Funding for the DPAC is provided by national institutions, in particular the institutions participating in the Gaia MultiLateral Agreement (MLA). The Gaia mission website is \url{https://www.cosmos.esa.int/gaia}. The Gaia archive website is \url{https://archives.esac.esa.int/gaia}.

Based on observations obtained with the Samuel Oschin Telescope 48-inch and the 60-inch Telescope at the Palomar Observatory as part of the Zwicky Transient Facility project. ZTF is supported by the National Science Foundation under Grants No. AST-1440341 and AST-2034437 and a collaboration including current partners Caltech, IPAC, the Weizmann Institute of Science, the Oskar Klein Center at Stockholm University, the University of Maryland, Deutsches Elektronen-Synchrotron and Humboldt University, the TANGO Consortium of Taiwan, the University of Wisconsin at Milwaukee, Trinity College Dublin, Lawrence Livermore National Laboratories, IN2P3, University of Warwick, Ruhr University Bochum, Northwestern University and former partners the University of Washington, Los Alamos National Laboratories, and Lawrence Berkeley National Laboratories. Operations are conducted by COO, IPAC, and UW.

\bibliographystyle{apsrev4-1}
\bibliography{main}

\appendix

\section{TESS orbital modulation}
\label{sec:appendix_tess}

\begin{figure}
    \centering
    \includegraphics[width=0.6\linewidth]{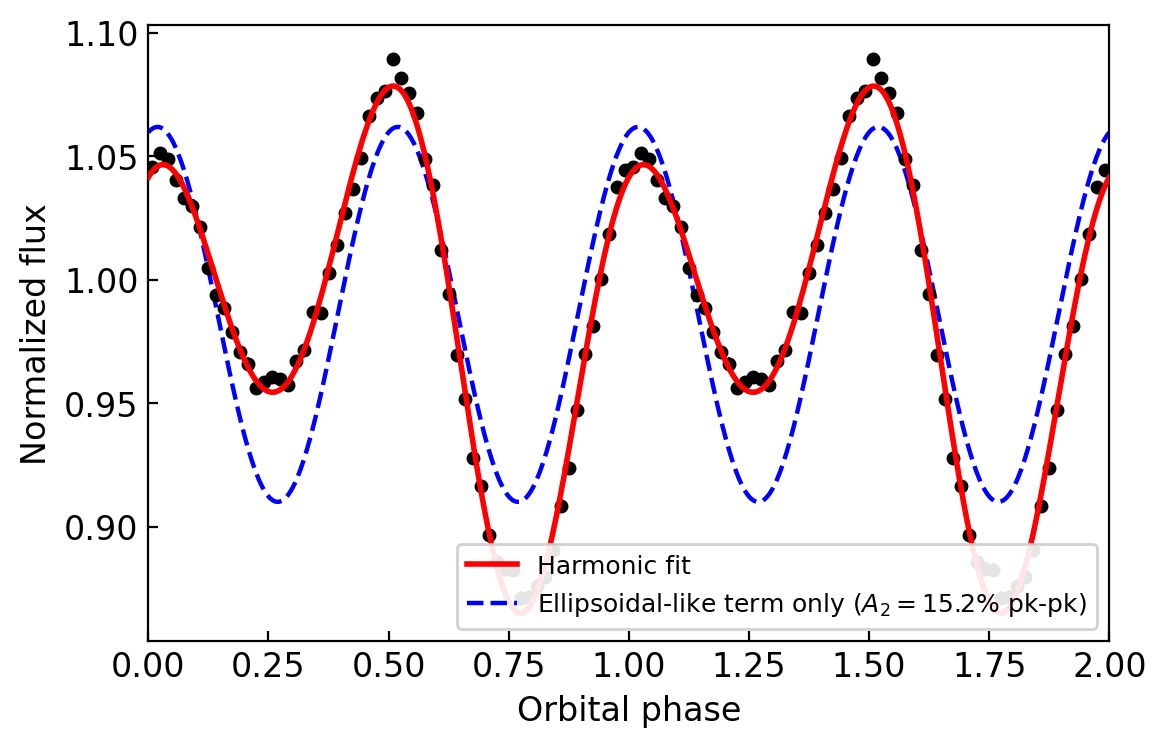}
    \caption{Phase-folded, flare-clipped TESS Sector 31 light curve of 1eRASS~J0426 (\textit{black}, two orbital cycles) with the three-term harmonic fit (\textit{red}) and the $\cos 4\pi\phi$ component (\textit{blue dashed}, $15.2\%$ peak-to-peak). The amplitude exceeds the maximum photospheric ellipsoidal signal for the adopted binary parameters by a factor of $\sim$3--6, so the modulation is dominated by accretion/propeller light and is not used as an inclination constraint (Appendix~\ref{sec:appendix_tess}).}
    \label{fig:ellipsoidal}
\end{figure}

A Lomb--Scargle periodogram of the flare-clipped TESS Sector~31 light curve (16515 good 2-min PDCSAP cadences, iteratively clipped against a running phase-binned median) peaks at $P = 3.308 \pm 0.005$~hr $= P_{\mathrm{orb}}/2$, the classic signature of an orbital-modulation-dominated light curve, confirming the orbital period used throughout. Fitting the phase-folded light curve with a three-term Fourier series gives a second-harmonic amplitude $A_2 = 0.0759 \pm 0.0004$ ($15.2$\% peak-to-peak) and an orbital-fundamental amplitude $A_1 = 0.041$ (Figure~\ref{fig:ellipsoidal}).

The $P_{\mathrm{orb}}/2$ modulation is \emph{not}, however, primarily photospheric ellipsoidal modulation. For the adopted binary parameters (Section~\ref{sec:joint}), the \citet{morris1985} ellipsoidal semi-amplitude,
\begin{equation}
    A_2^{\mathrm{ell}} \;=\; \frac{(15 + u)(1 + \tau)}{15 - 5u}\,q\,\left(\frac{R_{\mathrm{donor}}}{a}\right)^{\!3}\,\sin^2 i,
    \label{eq:morris}
\end{equation}
with standard limb- and gravity-darkening coefficients ($u \approx 0.55$, $\tau \approx 0.32$; \citealt{1967lucy}), is at most $A_2^{\mathrm{ell}} \approx 0.012$--$0.026$ across the allowed parameter range even at $i = 90^{\circ}$ -- a factor of $\sim$3--6 below the observed amplitude. The orbital modulation must therefore be dominated by non-photospheric light varying over the orbit (e.g.\ propeller/accretion light and irradiation, as also indicated by the large $A_1$ term, and as observed in AE~Aqr's own orbital light curve). We note that an earlier iteration of this analysis, which assumed a Roche-filling $\sim 0.7~M_\odot$ ZAMS-like donor, could attribute the full amplitude to the photosphere and thereby derived $i \approx 72^{\circ}$; with the much smaller measured donor radius and mass ratio, the photospheric signal cannot reach the observed amplitude at any inclination, and that quantitative constraint dissolves. The amplitude therefore cannot be converted into a quantitative inclination constraint, although the sheer strength of the modulation qualitatively favours a high inclination, since every plausible contributor scales as $\sin i$ or $\sin^2 i$. Eclipses provide no independent bound, because the WD contributes too little optical light (Section~\ref{sec:joint}).

\section{X-ray spectral analysis}
\label{sec:appendix_xray}

\begin{figure}
    \centering
    \includegraphics[width=0.55\linewidth]{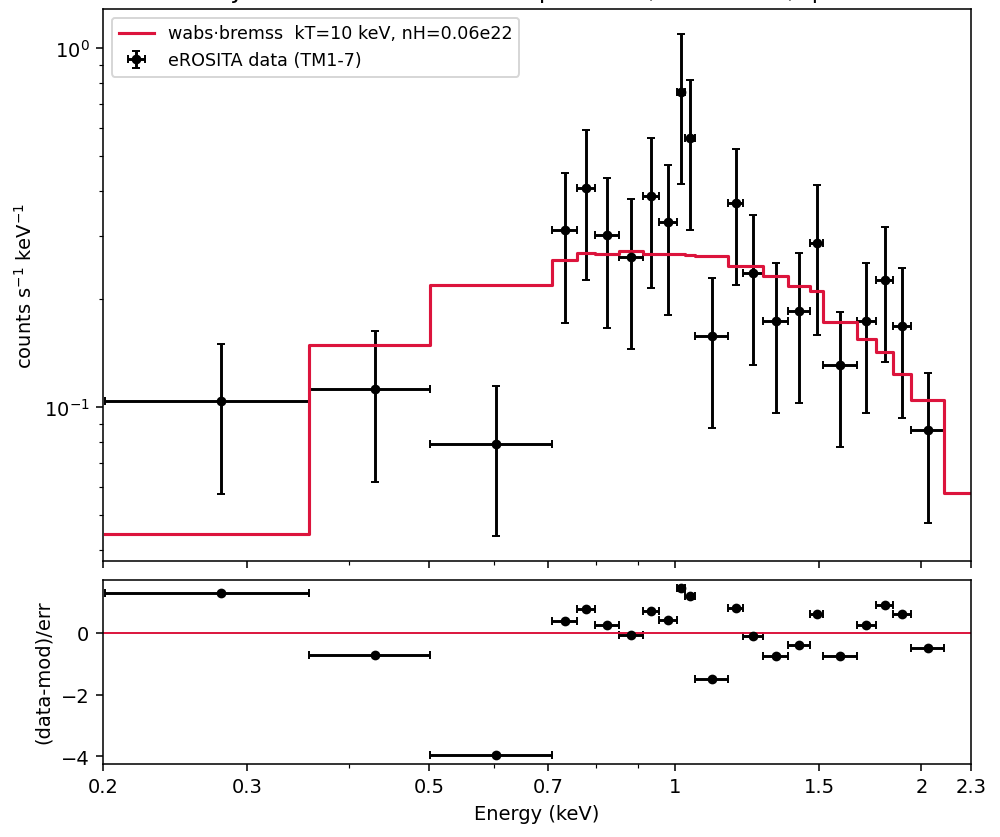}
    \caption{eROSITA eRASS1 0.2--2.3~keV spectrum of 1eRASS~J0426 (\textit{black}) with an absorbed thermal-bremsstrahlung model, $kT = 10$~keV and $N_H = 6\times 10^{20}$~cm$^{-2}$ (\textit{red}), and fit residuals (\textit{bottom}).}
    \label{fig:xray_spec}
\end{figure}

\begin{figure}
    \centering
    \includegraphics[width=0.6\linewidth]{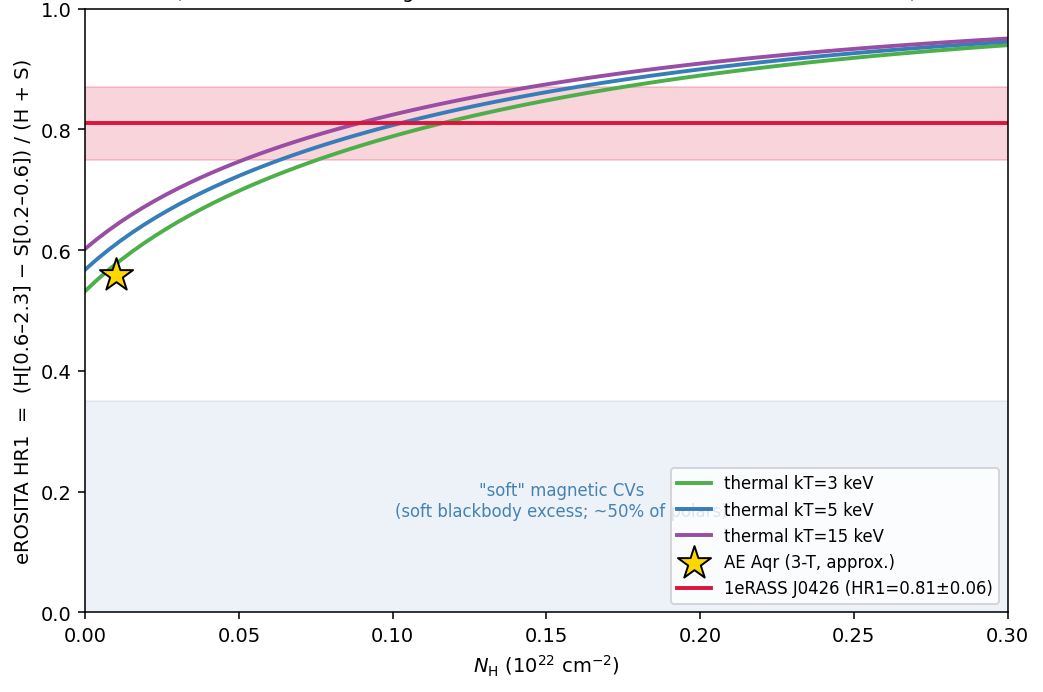}
    \caption{Hardness ratio $HR_1$ of 1eRASS~J0426 (\textit{red band}) compared with response-folded thermal-plasma models ($kT = 3, 5, 15$~keV) versus column $N_H$, and with AE~Aqr's three-temperature plasma \citep{2008mauche} folded through the same response (\textit{gold star}). The blue shaded region marks the soft, blackbody-excess regime occupied by $\sim$50\% of polars.}
    \label{fig:xray_hr}
\end{figure}

Figure~\ref{fig:xray_spec} shows the eROSITA eRASS1 spectrum with the absorbed thermal-bremsstrahlung fit (forward-folded through the ARF/RMF, Cash statistic): $kT \approx 10$~keV with $N_H \approx 6\times 10^{20}$~cm$^{-2}$. The temperature is bounded only from below ($kT \gtrsim 2$~keV at 1$\sigma$) because the soft 0.2--2.3~keV band contains no thermal cutoff to pin the high-temperature end; a partial-covering absorber is not statistically required. Figure~\ref{fig:xray_hr} compares the measured hardness ratio, $HR_1 = 0.81 \pm 0.06$ (with $S = 0.2$--$0.6$ and $H = 0.6$--$2.3$~keV), with thermal models of $kT = 3, 5, 15$~keV and with AE~Aqr's three-temperature plasma \citep{2008mauche}, all folded through this source's own eROSITA response so the comparison is self-consistent. 1eRASS~J0426 lies in the hard, low-$N_H$ region typical of magnetic CVs, far from the soft, blackbody-excess regime occupied by $\sim$50\% of polars; AE~Aqr folds to a slightly softer $HR_1$, consistent with both systems being in the same low-accretion regime.

\begin{figure}
    \centering
    \includegraphics[width=0.42\linewidth]{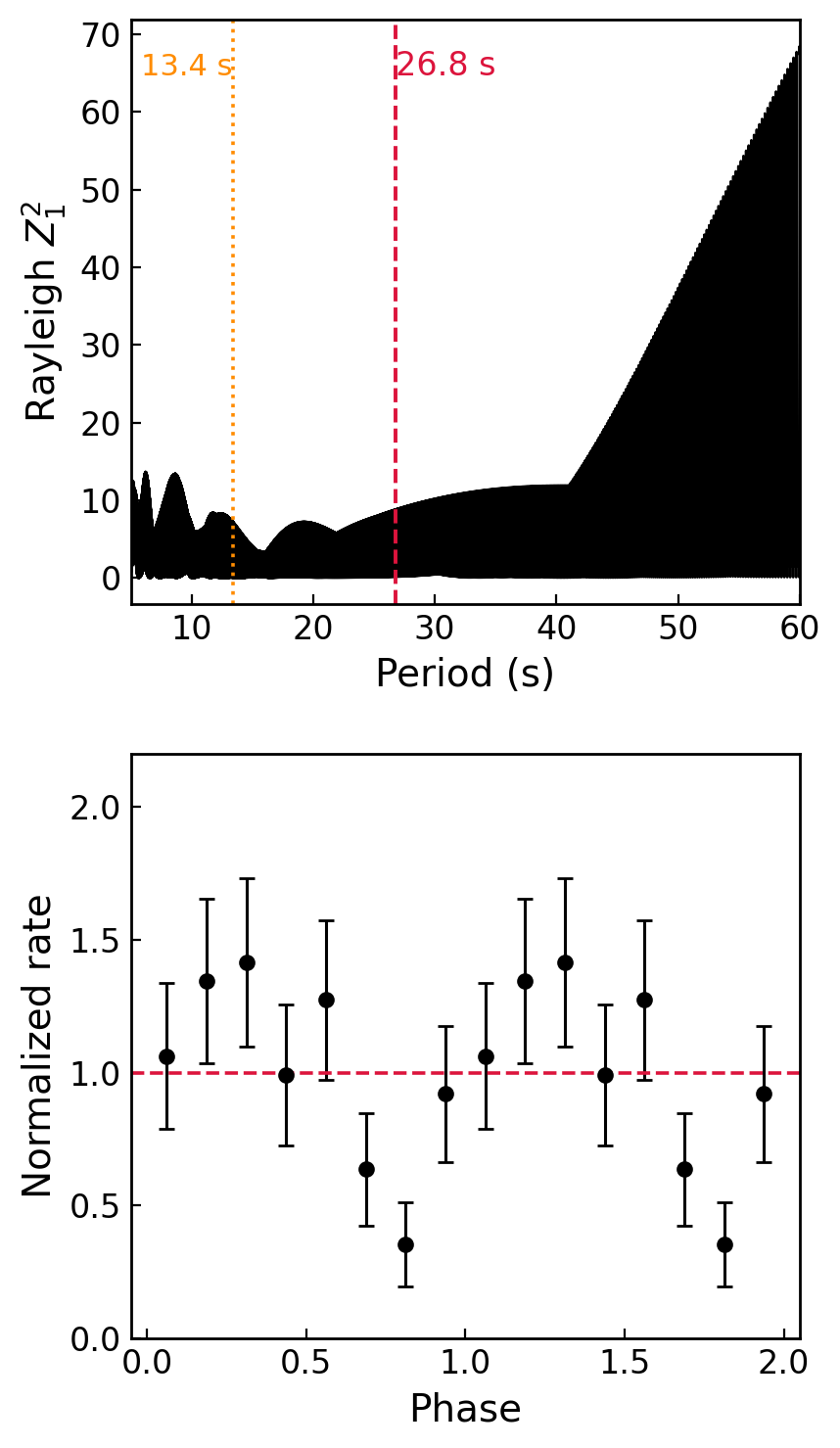}
    \caption{\textit{Top}: Rayleigh $Z^2_1$ periodogram of the barycentred eROSITA photons (0.2--2.3~keV, 113 photons) over 5--60~s, with the strongest CHIMERA candidate (26.8~s, \textit{red dashed}) and its first harmonic (13.4~s, \textit{orange dotted}) marked; the rising envelope toward longer periods is due to aliases of the 4-hr scan-visit cadence. A marginal ($\sim$2$\sigma$ window-corrected) enhancement appears at 26.8~s. \textit{Bottom}: barycentred photons folded at 26.8~s. The 99\% upper limit on the pulsed fraction is $\approx 40$\%.}
    \label{fig:xray_period}
\end{figure}

Figure~\ref{fig:xray_period} shows the X-ray periodicity search of Section~\ref{sec:xray}.

\section{Donor temperature from the low-state spectrum}
\label{sec:appendix_donor}

To test the SED-derived donor temperature, we compare the low-state DBSP spectrum directly with stellar templates. Each of the 38 low-state spectra is shifted to the donor rest frame using the RV solution of Section~\ref{sec:rv} and median-combined. BT-Settl templates \citep{2011allard} of $\log g = 4.5$ and solar metallicity, spanning $T_{\mathrm{eff}} = 3800$--$4600$~K, are smoothed to the instrumental resolution, reddened by $E(B-V)=0.058$, and fitted to the observed spectrum with a single scale factor over 4500--9200~\AA{}, masking emission lines, telluric bands, and the dichroic crossover.

\begin{figure}
    \centering
    \includegraphics[width=0.75\linewidth]{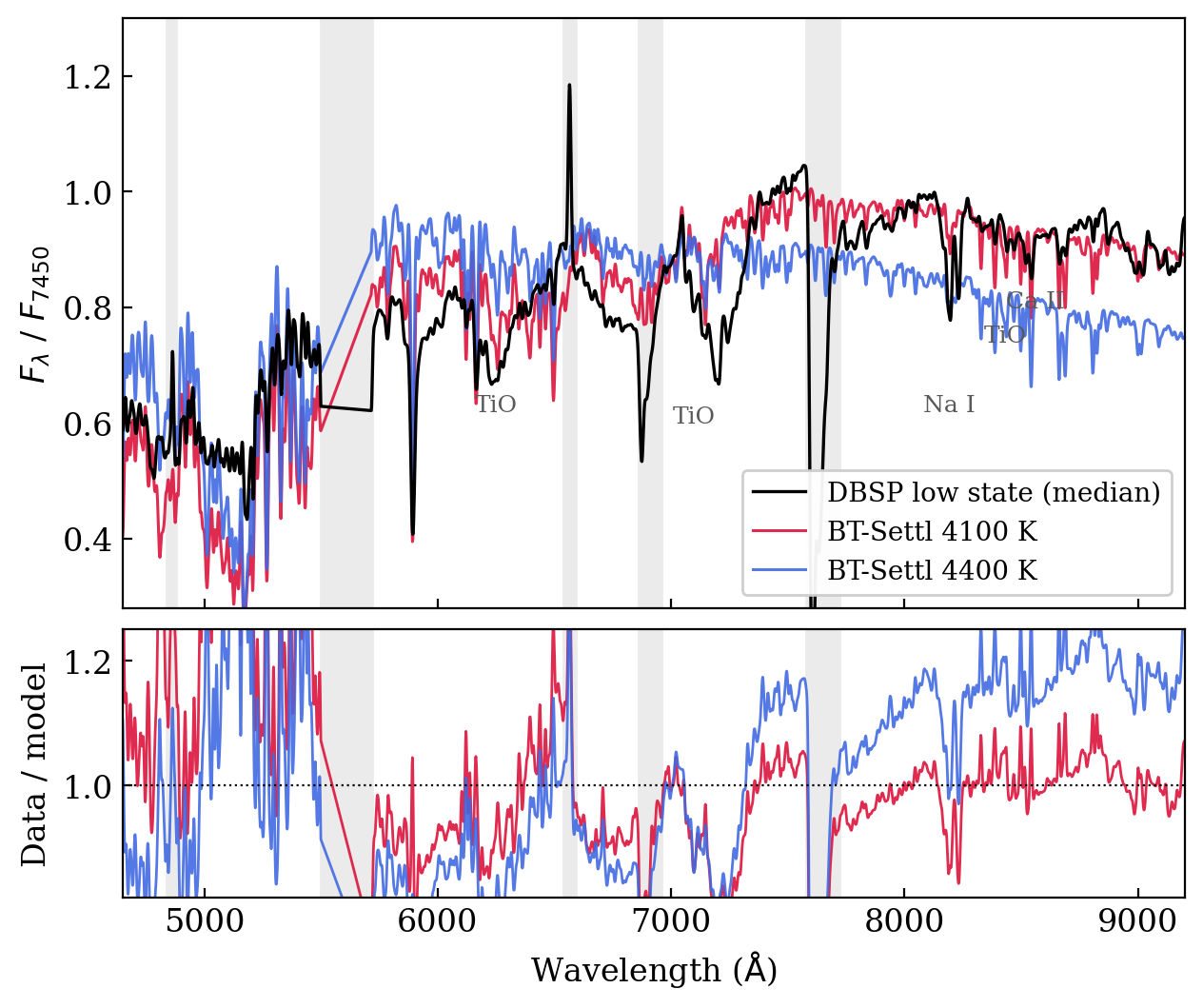}
    \caption{\textit{Top}: median low-state DBSP spectrum in the donor rest frame (\textit{black}) compared with BT-Settl templates at 4100~K (\textit{red}) and 4400~K (\textit{blue}), each scaled by a single factor fitted over 4500--9200~\AA{}; grey bands mark masked emission-line, telluric, and dichroic-crossover regions. \textit{Bottom}: data/model ratio. The 4400~K template systematically under-predicts the red flux, while 4100~K tracks the data; the observed TiO band depths also require the cooler temperature.}
    \label{fig:donor_temp}
\end{figure}

Figure~\ref{fig:donor_temp} shows the result. The $\chi^2$ minimum lies at $T_{\mathrm{eff}} \approx 4100$--$4200$~K, and $T_{\mathrm{eff}} \geq 4300$~K is strongly disfavoured: a 4400~K template under-predicts the observed flux redward of $\sim$7300~\AA{} by 15--20\%, and the observed TiO bands (6250, 6870, 7100~\AA) are deeper than even the 4100~K template predicts. The 2MASS colours ($J-K_s = 0.82$) independently indicate $T_{\mathrm{eff}} \approx 3900$--$4100$~K. We therefore adopt $T_{\mathrm{donor}} = 4100 \pm 150$~K. The hotter SED-only value (4345~K; Section~\ref{sec:sed}) is naturally explained by contamination of the state-averaged optical survey photometry by blue accretion light, the same systematic that produces the SkyMapper $u/g$ excess.

\section{Alternative Zeeman field solution at 33 MG}
\label{sec:appendix_zeeman33}

\begin{figure*}
    \centering
    \includegraphics[width=0.95\linewidth]{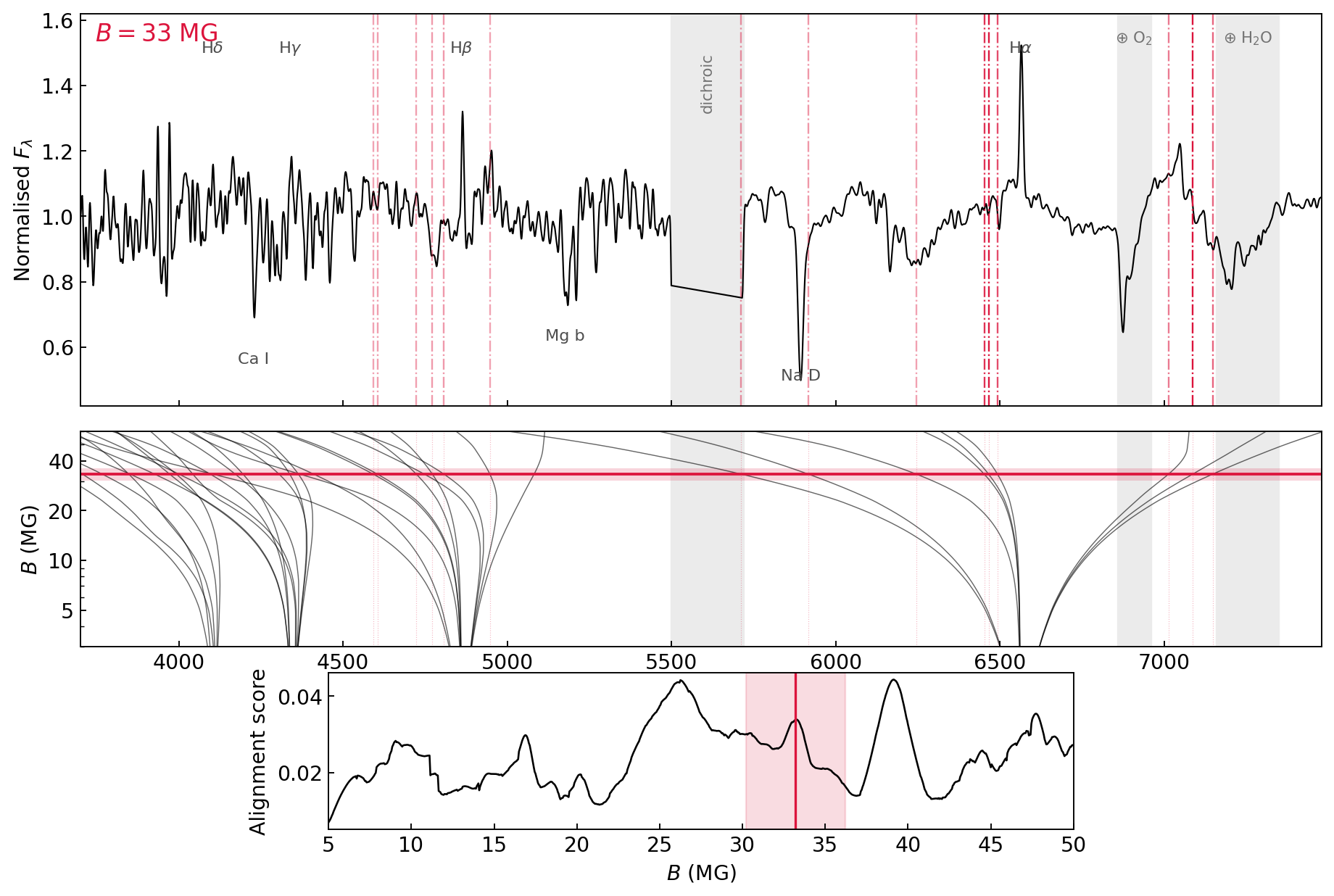}
    \caption{Same as Figure~\ref{fig:zeeman}, but for the alternative field solution $B = 33$~MG, the secondary maximum of the alignment score.}
    \label{fig:zeeman33}
\end{figure*}

The alignment score of Section~\ref{sec:zeeman} shows a secondary maximum at $B \approx 33$~MG that is nearly as strong as the adopted 26~MG optimum. Figure~\ref{fig:zeeman33} shows the corresponding component identifications: at 33~MG a different subset of predicted Balmer components aligns with the detected low-state absorption features, most notably around H$\beta$. Because the alignment statistic alone cannot formally exclude this solution, we regard $B_{\mathrm{WD}} \approx 33$~MG as a viable alternative to the adopted value; either field places the WD in the same transitional regime between IPs and polars, and the propeller conclusion of Section~\ref{sec:propellermode} is unchanged (at fixed $\dot M$, a higher field only increases $P_{\mathrm{eq}}$).

\section{CHIMERA pulsation searches on additional nights}
\label{sec:appendix_chimera}

\begin{figure*}
    \centering
    \includegraphics[width=0.9\linewidth]{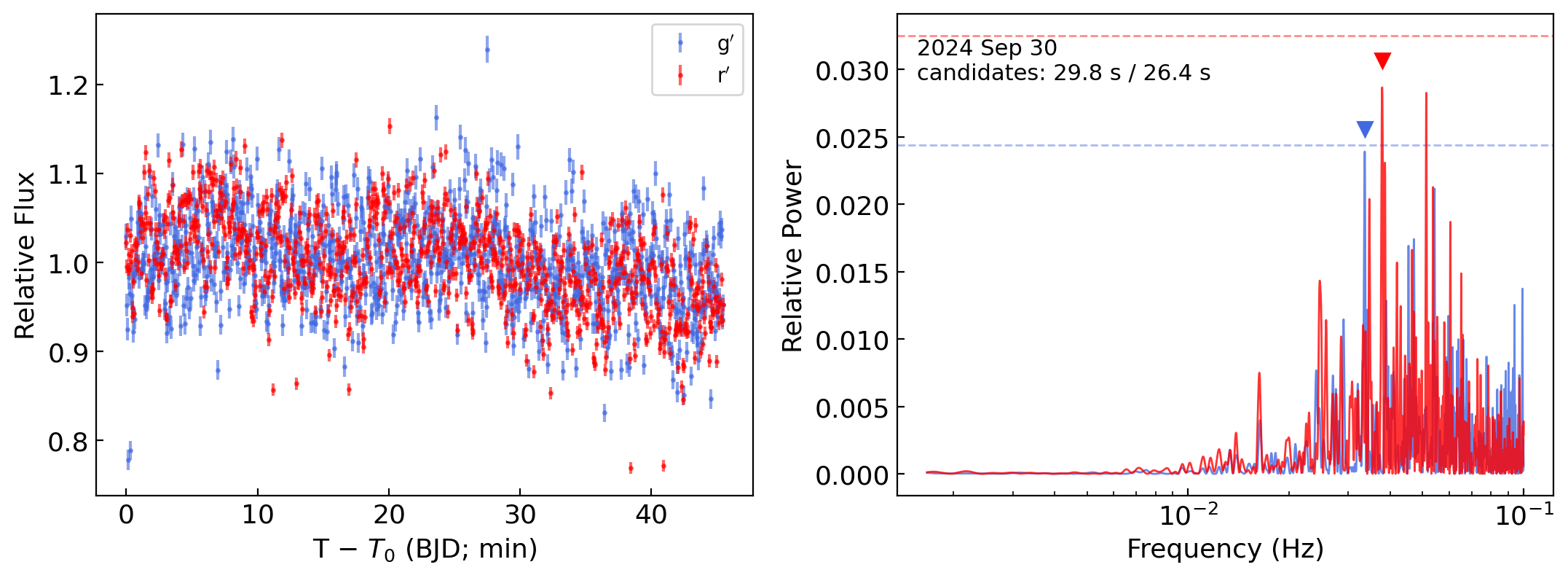}
    \caption{CHIMERA photometry of 1eRASS~J0426 from 2024 September 30. \textit{Left}: $g'$ (\textit{blue}) and $r'$ (\textit{red}) light curves. \textit{Right}: Lomb--Scargle periodograms of the boxcar-detrended light curves; triangles mark the strongest peaks (29.8~s in $g'$, 26.4~s in $r'$), and the dashed lines the 10$\sigma$ levels.}
    \label{fig:chimera_0930}
\end{figure*}

\begin{figure*}
    \centering
    \includegraphics[width=0.9\linewidth]{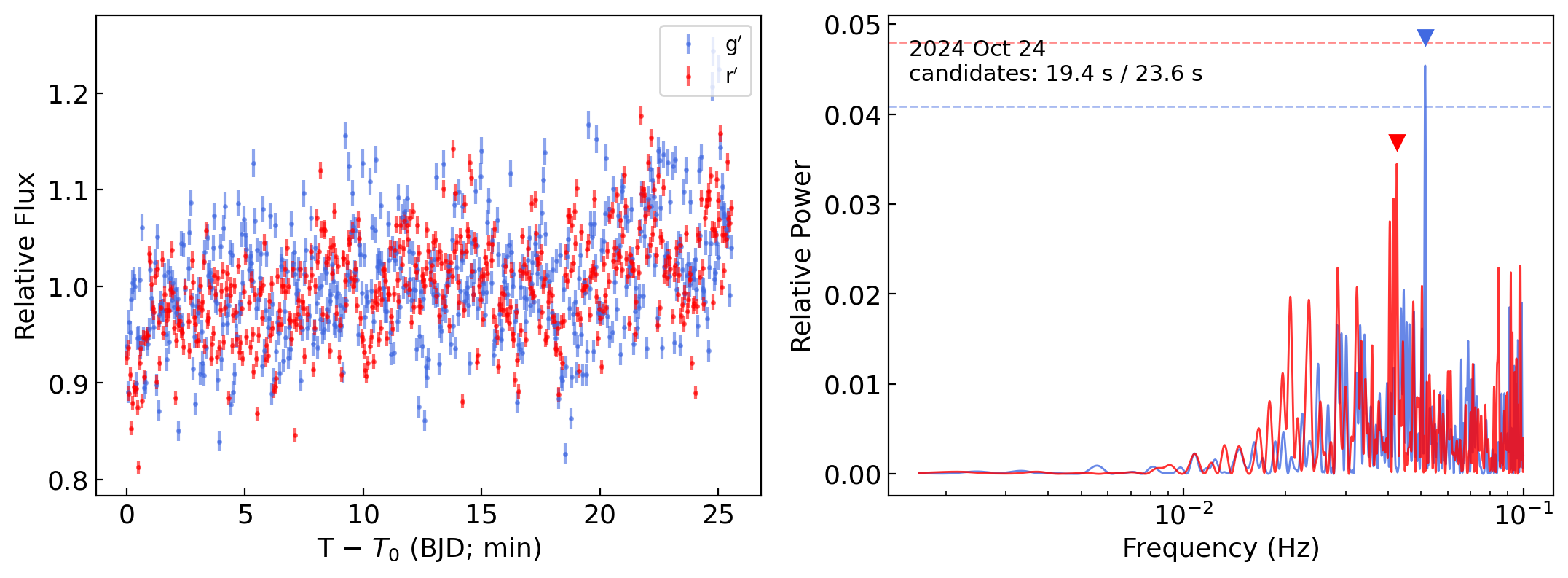}
    \caption{Same as Figure~\ref{fig:chimera_0930}, for 2024 October 24. The strongest peaks lie at 19.4~s ($g'$) and 23.6~s ($r'$).}
    \label{fig:chimera_1024}
\end{figure*}

Figures~\ref{fig:chimera_0930} and \ref{fig:chimera_1024} show the CHIMERA light curves and Lomb--Scargle periodograms for the 2024 September 30 and October 24 runs, constructed identically to Figure~\ref{fig:chimera}: the light curves are divided by an 11-point boxcar smoothing to remove the flaring envelope, and the periodogram is computed over periods of 10~s--10~min. On September 30 the strongest peaks lie at 29.8~s ($g'$) and 26.4~s ($r'$); on October 24 they lie at 19.4~s ($g'$) and 23.6~s ($r'$). In each case several nearly-equal-power aliases surround the strongest peak, reflecting the pulse-amplitude and phase wander during flaring activity; these night-to-night variations motivate the conservative $P_{\mathrm{spin}} < 60$~s constraint adopted in Section~\ref{sec:chimera}.

\section{Lick/Shane spectroscopy}
\label{sec:appendix_kast}

\begin{figure}
    \centering
    \includegraphics[width=0.8\linewidth]{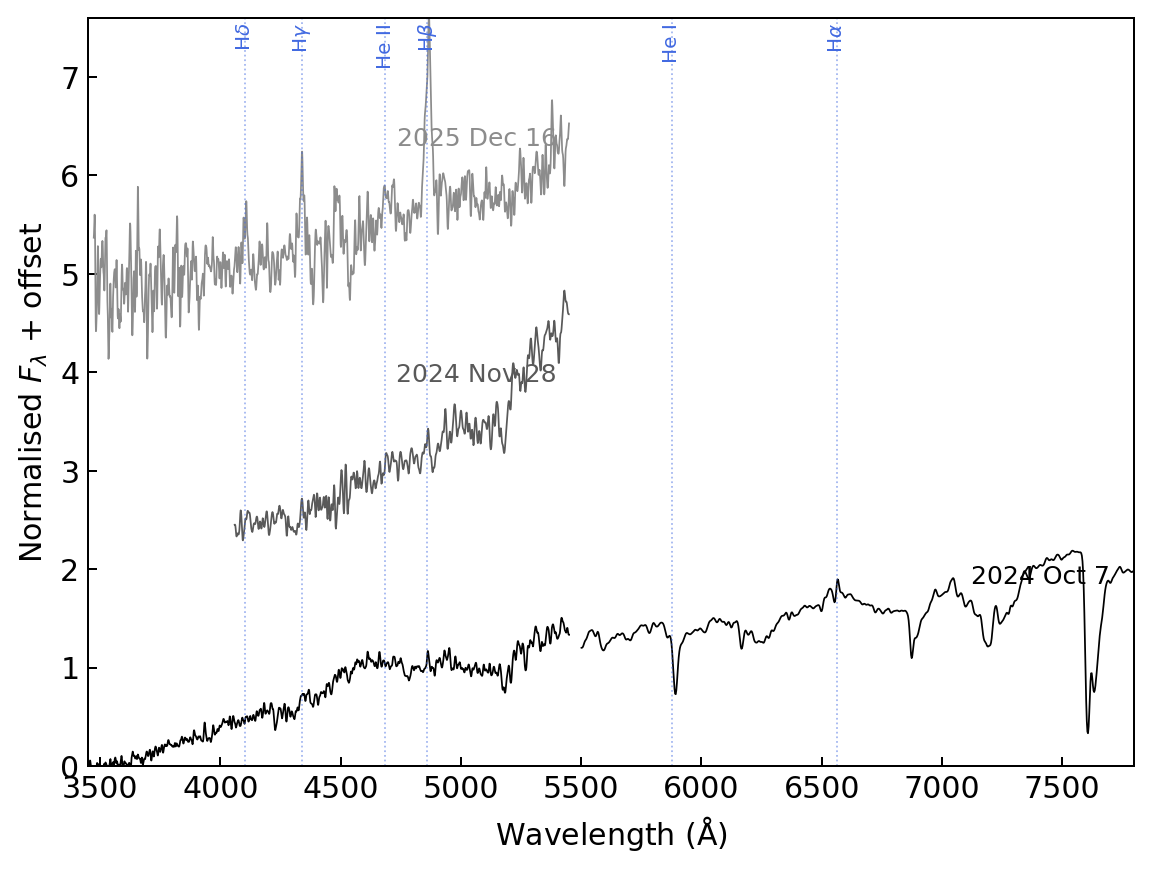}
    \caption{Lick/Shane Kast spectra of 1eRASS~J0426 from 2024 October 7 (blue + red arms), 2024 November 28 (blue only), and 2025 December 16 (blue only), normalised near 5000~\AA{} and vertically offset. Balmer and He lines are marked. The 2025 December spectrum shows the strongest Balmer emission.}
    \label{fig:kast}
\end{figure}

1eRASS~J0426 was observed with the Kast spectrograph on the Shane 3-m telescope at Lick Observatory on three additional epochs: 2024 October 7 (4$\times$blue and 4$\times$red exposures; MJD 60590.41--60590.44), 2024 November 28 (2$\times$blue; MJD 60642.42), and 2025 December 16 (1$\times$blue; MJD 61025.28). The median-combined spectra are shown in Figure~\ref{fig:kast}. All three epochs show a donor-dominated red continuum. The 2024 October and November spectra show only weak Balmer emission, consistent with the low photometric state at those epochs (Figure~\ref{fig:ztf}), while the 2025 December spectrum shows stronger Balmer emission (H$\beta$, H$\gamma$, H$\delta$), suggesting the system was in (or entering) a more active state at that epoch.

\section{MCMC corner plots}
\label{sec:appendix_corner}

\begin{figure}
    \centering
    \includegraphics[width=0.7\linewidth]{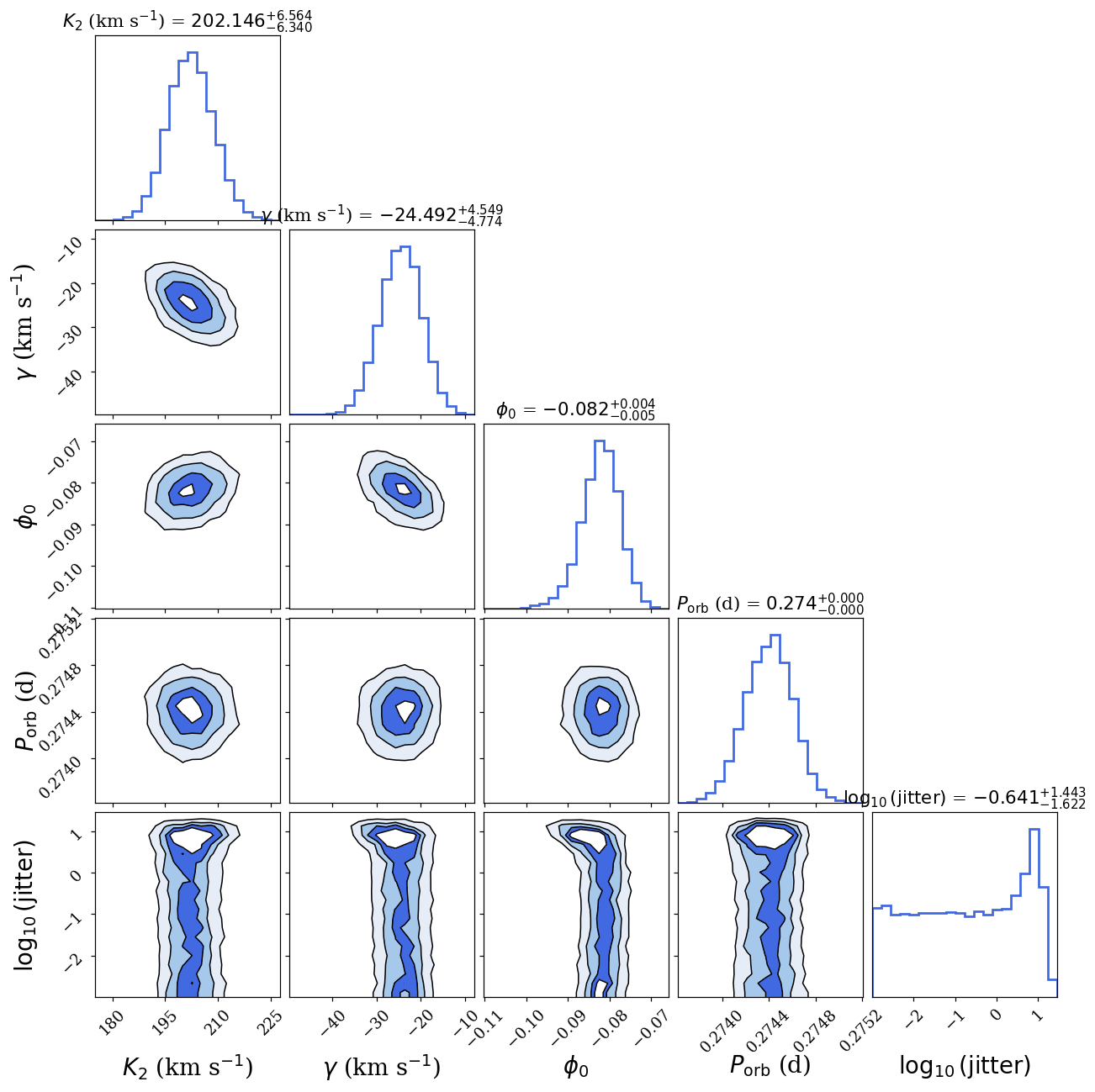}
    \caption{Posteriors of the RV fit (Eq.~\ref{eq:rv}). $K_2$, $\gamma$, and $\phi_0$ are well constrained; the per-point jitter is consistent with zero.}
    \label{fig:rv_corner}
\end{figure}

\begin{figure}
    \centering
    \includegraphics[width=0.7\linewidth]{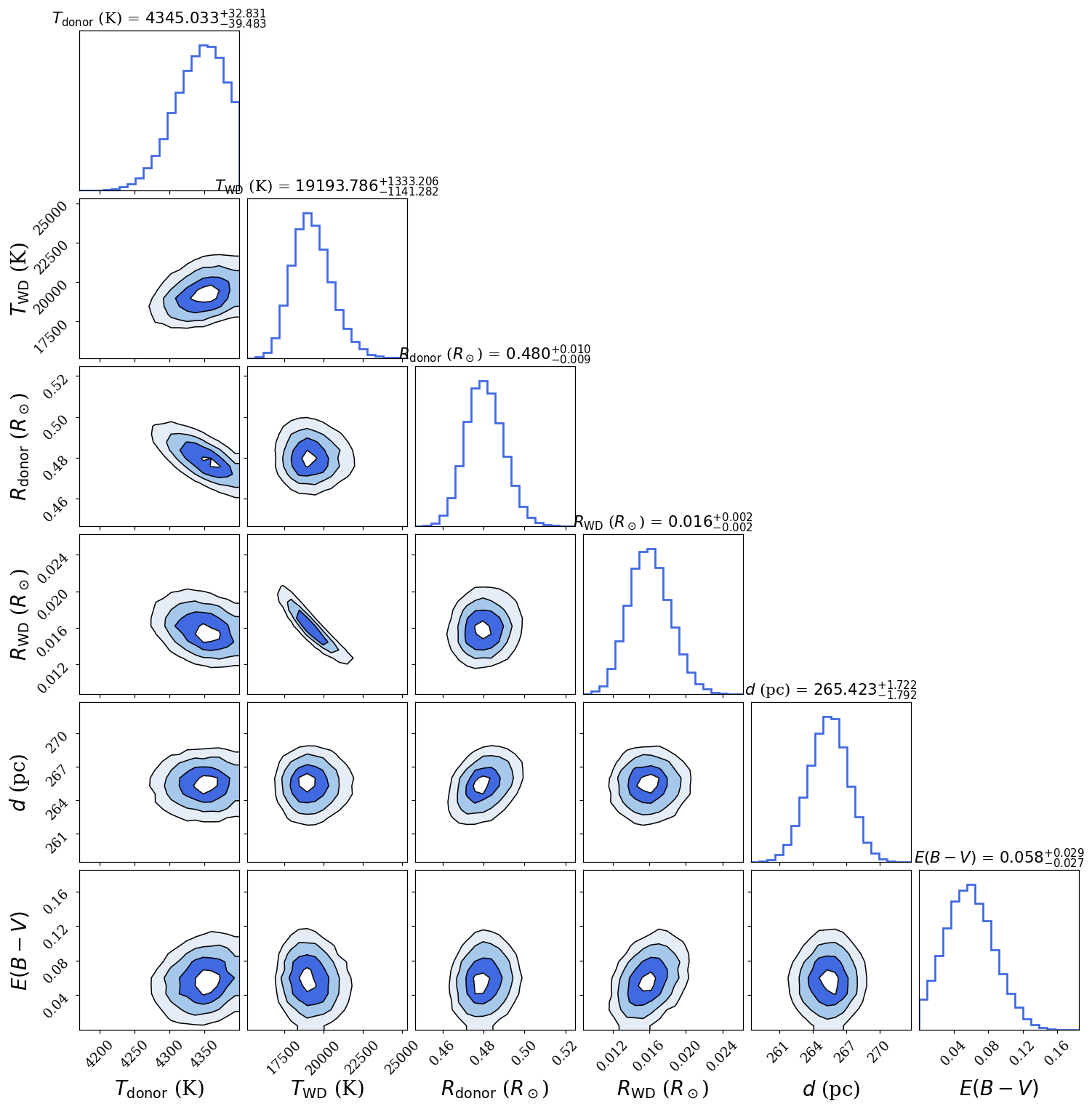}
    \caption{Posteriors of the SED fit. The donor temperature and radius are tightly determined by the Gaia, 2MASS, and WISE photometry; $T_{\mathrm{WD}}$ and $R_{\mathrm{WD}}$ should be regarded as upper limits given possible non-photospheric UV contributions.}
    \label{fig:sed_corner}
\end{figure}

Figures~\ref{fig:rv_corner} and \ref{fig:sed_corner} show the marginalised posteriors of the radial-velocity (Section~\ref{sec:rv}) and SED (Section~\ref{sec:sed}) fits. The RV posterior includes a per-point jitter and a tight Gaussian prior on $P_{\mathrm{orb}}$ centred on the TESS value (the period posterior reproduces the prior). The SED posterior includes a parallax-pinned distance prior and an SFD-anchored prior on $E(B-V)$.

\end{document}